\documentclass[%
 reprint,
 amsmath,amssymb,
 aps,
]{revtex4-2}
\usepackage{float}
\usepackage{xcolor}
\usepackage{hyperref}
\usepackage{graphicx}
\usepackage{dcolumn}
\usepackage{bm}

\begin{document}

\preprint{APS/123-QED}

\title{Superconductivity and strain-enhanced phase stability of Janus tungsten chalcogenide hydride monolayers}

\author{Jakkapat Seeyangnok$^{1,2}$}
 \email{jakkapatjtp@gmail.com} 
\author{Udomsilp Pinsook$^{2}$}%
 \email{Udomsilp.P@Chula.ac.th}

\author{Graeme J Ackland$^{1}$}
 \email{gjackland@ed.ac.uk} 
\affiliation{$^{1}$Centre for Science at Extreme Conditions, School of Physics and Astronomy, University of Edinburgh, Edinburgh, United Kingdom\\$^{2}$Department of Physics, Faculty of Science, Chulalongkorn University, Bangkok, Thailand.}%


\date{\today}

\begin{abstract}
    Janus transition metal-dichalcogenide (JTMD) materials have attracted a great deal of attention due to their remarkable physical properties arising from the two-dimensional geometry and the breakdown of the out-of-plane symmetry. Using first-principles density functional theory, we investigated the phase stability, strain-enhanced phase stability, and superconductivity of Janus WSeH and WSH. In addition, we investigated the contribution of the phonon linewidths from the phonon energy spectrum responsible for the superconductivity, and the electron-phonon coupling as a function of phonon wave vectors and modes. Previous work has examined hexagonal 2H and tetragonal 1T structures, but we found that neither is a ground state structure. The metastable 2H phase of WSeH is dynamically stable with T$_{c}$ $\approx$ 11.60K, similar to WSH. Compressive biaxial strain - the 2D equivalent of pressure - can stabilize the 1T structures of WSeH and WSH with T$_{c}$ $\approx$ 9.23K and 10.52K, respectively. 
\end{abstract}

\maketitle


	\section{Introduction}\label{sec1}
     Since the discovery of graphene in 2004 \cite{novoselov2004electric}, two-dimensional (2D) materials have attracted attention due to their rich physical properties and various possible applications in nanomaterials and condensed matter physics. At the same time, superconducting hydrides have set new records for high critical temperatures. These materials are inspired by the prediction of room temperature superconductivity at extreme pressures in metallic hydrogen and its alloys proposed by Ashcroft \cite{ashcroft1968metallic,ashcroft2004hydrogen,mcmahon2011high,liu2017potential,zurek2009little}. It is attractive to consider combining these classes of materials. Therefore, a new form of 2D hydrides has been investigated to obtain possible high-temperature superconductivity without requiring huge pressure along with many interesting physical properties resulting from the 2D geometry. The very first candidate for 2D binary hydrides was hydrogenated graphene which could become a conventional superconductor with $T_{c}$ above 90K \cite{sofo2007graphane}. Furthermore, 2D ternary hydrides of hydrogenated MgB$_{2}$ monolayer (ML) was theoretically proposed with $T_{c}$ = 67K \cite{savini2010first}, and hydrogenated HPC$_{3}$ was shown to be a conventional superconductor of predicted $T_{c}$ = 31K \cite{li2022phonon}. Multilayer graphene itself displays remarkable "magic-angle" superconductivity\cite{cao2018unconventional}, adding further potential for breakthroughs.
	
    Two-dimensional transition metal dichalcogenides (2D-TMDs) have an out-of-plane symmetric layered crystal structure composed of a monolayer (ML) of a transition metal atom, with the lower and upper facets containing chalcogenide atoms of the same chemical element. Much attention has been paid to TMDs due to their flat electronic band structures \cite{calandra2018phonon, rostami2018helical, tresca2019charge, nakata2021robust}, and induced electronic flat bands in TMDs through heterostructures and twistronics \cite{zhang2020flat, vitale2021flat, kuang2022flat, rademaker2022spin, huang2023recent}. However, 2D Janus transition metal dichalcogenides (2D-JTMDs) are layered out-of-plane asymmetric crystal structures where the lower and upper facets contain different chalcogenide atoms. 2D-JTMDs exhibit interesting and tunable electronic, optical, and mechanical properties due to structural asymmetry between the layers\cite{tang20222d,varjovi2021janus,zhang2022janus,angeli2022twistronics,maghirang2019predicting,he2018two,yeh2020computational,yin2021recent,li2023structure}. Although 2D-JTMDs may not exist naturally, they have been successfully synthesized, starting with Janus graphene in 2013 \cite{zhang2013janus}. Many more 2D-JTMDs are now known, for example, MoSSe \cite{trivedi2020room,lu2017janus}, WSSe \cite{trivedi2020room}, and PtSSe\cite{sant2020synthesis}. Successful fabrication of 2D-JTMDs involves a selective epitaxy atomic replacement (SEAR) process using hydrogen (H$_{2}$) plasma \cite{trivedi2020room,tang20222d}. 
    
    2D-TMDs can form many different phases, depending on the position of the upper and lower facets. The known phases are the 2H, 1T, 1T$^{'}$ and 2a$\times$2a phases \cite{zhuang2017doping}. Thus, it is natural to expect that 2D-JTMDs will also exhibit different phases. Recently, a Janus 2H-MoSH ML has been synthesized using the SEAR method by replacing the S atoms on the top facet with hydrogen atoms \cite{lu2017janus}. Janus MoSH ML has recently been predicted to be a superconductor with T$_{c}$ = 27K for the 2H phase \cite{liu2022two} and T$_{c}$ = 25K for the 1T phase \cite{ku2023ab}. Furthermore, the very first calculations on 2H-WSH have discovered that it is dynamically stable, has a metallic phase, and becomes a superconductor with T$_{c}$ above 12K \cite{seeyangnok2024superconductivity}. Later, the phase stability and superconductivity of the 2H phase of WSH have been confirmed by independent and subsequent research \cite{gan2024hydrogenation,fu2024superconductivity}. Recently, there has also been a report on the superconductivity of the TiSH monolayer \cite{TiSH}.
    
    In general, 2D materials are prepared and synthesized on suitable substrates \cite{lu2017janus, wan2021synthesis}, which inevitably induce strain on the system. The effect of strain on physical properties and applications in 2D materials has been intensively investigated \cite{postorino2020strain,yang2021strain,blundo2021strain}, with extensive investigations on the transition metal dichalcogenides \cite{sharma2014strain,rostami2015theory,shen2016strain,lee2017strain,wei2017straintronics,khatibi2018impact,bendavid2022strain}. For example, strain can stabilize the 1T phase of MoSH \cite{ku2023ab}. In addition, strain significantly affects the superconducting properties, for example, T$_{c}$ of the hydrogenated MgB$_{2}$ monolayer (ML) and hydrogenated HPC$_{3}$ can be increased up to 100K \cite{sofo2007graphane,savini2010first} by applying biaxial strain 5\% and 3\%, respectively. 
    
    Phonon-mediated superconductors \cite{bardeen1957microscopic} can be explained by the electron-phonon interaction based on the Migdal-Eliashberg (ME) theory \cite{frohlich1950theory,migdal1958interaction,eliashberg1960interactions,nambu1960quasi}. In ME theory, the electron-phonon coupling (EPC) acts as a glue for the pairing process of Cooper pairs, and also renormalizes the phonon and electron quasiparticles. An important implication of the renormalized quasiparticles is that it leads to the so-called phonon linewidth, which is connected to the imaginary part of the phonon self-energy of the Feynman one-bubble diagram \cite{allen1972neutron}. The phonon linewidth can be measured by inelastic neutron or X-ray scattering experiments, and provides important parameters in the study of phonon-involved phenomena such as lattice vibrations, scattering rates, phonon damping, and phonon modes for Cooper pairing. Therefore, it is very important to understand the phonon linewidth for a detailed analysis. It should be noted that standard first-principles lattice dynamics uses the harmonic approximation, but any comparison to the experiment may need to take into account additional effects of anharmonicity, which can be used to further renormalize phonon frequencies to incorporate into the extended Eliashberg theory of superconductivity \cite{errea2013first}. This is often the case for hydrogen rich materials wherein zero point motion can stabilize imaginary phonons. However, we have tested that anharmonic effects can be safely neglected by using the frozen-phonon method. Atoms are displaced simultaneously along the eigenvectors of the  flexural acoustic phonon modes responsible for the electron-phonon coupling. The curvature of the energy as a function of the squared displacement shows a linear behavior well above the corresponding zero-point phonon energy. The results indicate that the electron-phonon coupling driving the superconducting state is within the harmonic approximation regime.
	
	In this paper, we systematically examined the 2H and 1T phases of tungsten-selenium hydride (WSeH) and tungsten-sulfur hydride (WSH). We started with a discussion of the electronic structure. Then, we examine the phase stability and the strain-enhanced phase stability. In the following sections, we consider the superconductivity and the corresponding phonon linewidths that are responsible for the superconductivity. These should give insightful information on the superconductivity of this class of 2D materials.
	\section{Methodology}
    The calculation was performed based on the DFT implemented in QE \cite{giannozzi2009quantum,giannozzi2017advanced}. The crystal structures were created using VESTA \cite{momma2011vesta} with the space group of $P3m1$. The optimized structures were achieved using the BFGS method \cite{BFGS,liu1989limited}, fully relaxing with a force threshold of $10^{-5}$ eV/\AA. The vacuum thickness was set to be 20\AA\  with the Coulomb truncation along the z-axis \cite{sohier2017density,sohier2017breakdown} to obtain free-standing monolayers. The GGA-PBE pseudopotentials \cite{hamann2013optimized,schlipf2015optimization,perdew1996generalized} were used for the exchange-correlation energy functional with the wavefunction and charge density cutoffs of 80 Ry and 320 Ry, respectively, and a Monkhorst-Pack grid k-mesh \cite{monkhorst1976special} of 24$\times$24$\times$1 k-point grids with the smearing of 0.02Ry on the Fermi surface \cite{marzari1999thermal}.
	
	The electron-phonon interaction was calculated using DFPT with dense k-point grids of 48$\times$48$\times$1 which contains $\vec{k}+\vec{q}$ of 24$\times$24$\times$1 k-point grids and 12$\times$12$\times$1 q-point grids. In this study, we imposed the interatomic force constants (IFC) with Born-Huang invariance conditions and Haung conditions \cite{born1955dynamical,lin2022general} for the rotational invariance of the potential energy and the vanishing stress, respectively. 
 
 From the corrected IFCs, we calculated the phonon dispersion and phonon density of states and the phonon linewidth, $\gamma_{\boldsymbol{q}\nu}$, for each phonon band $\nu$ at wavevector $q$. 
 
A useful quantity calculated by QuantumEspresso is the sum of the electron-phonon coupling for each transition between electronic bands $n,m$ and each phonon ${\bf q}\nu$.
\begin{equation}  \label{eq:bandsum}
T_{nm{\bf q}\nu} =
\sum_{\boldsymbol{k}}|g_{\boldsymbol{k}+\boldsymbol{q},\boldsymbol{k}}^{\boldsymbol{q}\nu,mn}|^{2}\delta(\epsilon_{\boldsymbol{k}+\boldsymbol{q},m}-\epsilon_{F})\delta(\epsilon_{\boldsymbol{k},n}-\epsilon_{F})
\end{equation}
 where $g$ is the first order derivative of the Kohn-Sham potential with respect to the phonon displacements.
From this we get the phonon lifetime
	\begin{equation}  \label{gammaphononlinewidths}
\gamma_{\boldsymbol{q}\nu} = 2\pi\omega_{\boldsymbol{q}\nu}\sum_{nm} T_{nm{\bf q}\nu} 
	\end{equation}

     and the electron-phonon coupling, $\lambda_{\boldsymbol{q}\nu}$, associated with the phonon wavevector $\boldsymbol{q}$ and the phonon mode of $\mu$, as
    \begin{equation}\label{eqn:lambda_qv}
        \lambda_{\boldsymbol{q}\nu} = \frac{1}{N(\epsilon_{\textbf{F}})\omega_{\boldsymbol{q}\nu}}\sum_{nm}T_{nm{\bf q}\nu}
    \end{equation}
	where the double delta functions can be expressed in the integral form \cite{wierzbowska2005origins}.
Also, $\gamma_{\boldsymbol{q}\nu}$ and $  \lambda_{\boldsymbol{q}\nu} $  are calculated on a grid of {\bf $q$} points.
 These are then interpolated to evaluate the the Eliashberg spectral function, $\alpha^{2} F(\omega)$, as
\begin{equation}
	\alpha^{2}F(\omega)=\frac{1}{2N(\epsilon_{F})}\sum_{\boldsymbol{q}\nu}\delta(\omega-\omega_{\boldsymbol{q}\nu})   \frac{\gamma_{\boldsymbol{q}\nu}}{\omega_{\boldsymbol{q}\nu}}
\end{equation}
	The superconducting transition temperature (T$_{C}$) was calculated using the Allen-Dynes formula \cite{allen1975transition} as
	\begin{equation} \label{allencomputetc}
		T_{c} = \frac{\omega_{\text{ln}}}{1.20}\exp\left(-\frac{1.04(1+\lambda)}{\lambda-\mu^{*}(1+0.62\lambda)}\right)
    \end{equation}
	where the electron-phonon coupling constant $\lambda$ can be calculated from the Eliashberg spectral function by using 
	\begin{equation} \label{computelambdaformulas}
		\lambda =2\int^{\omega_{\text{max}}}_{0}d\Omega\left(\frac{\alpha^{2}F(\Omega)}{\Omega}\right),
	\end{equation}
	and the logarithmic average phonon energy can be computed by using 
	\begin{equation} \label{computeomegaln}
		\omega_{\text{ln}} = \exp\left(\frac{2}{\lambda}\int^{\infty}_{0}d\Omega\text{ln}(\Omega)\left(\frac{\alpha^{2}F(\Omega)}{\Omega}\right)\right).
	\end{equation}

	Finally, the Morel-Anderson pseudopotential was set to $\mu^* = 0.1$ for practical purposes; however, $T_c$ has some dependence on $\mu^*$  as shown in Appendix D.

	For normal in-plane biaxial strain, we applied the 2D strain tensor $\varepsilon_{ij}$ 
	where we used the same strain component for $\varepsilon_{xx}$ and $\varepsilon_{yy}$. In terms of computation, we first applied the strain tensor on the simulation cells, i.e. imposing the deformation onto the lattice constants. Then, the strained structures were fully relaxed to obtain the fully relaxed crystal structures.

	\section{Structural Properties}
	\subsection{Crystal structures}
 
	\begin{figure}[h]
		\centering
		\includegraphics[width=9cm]{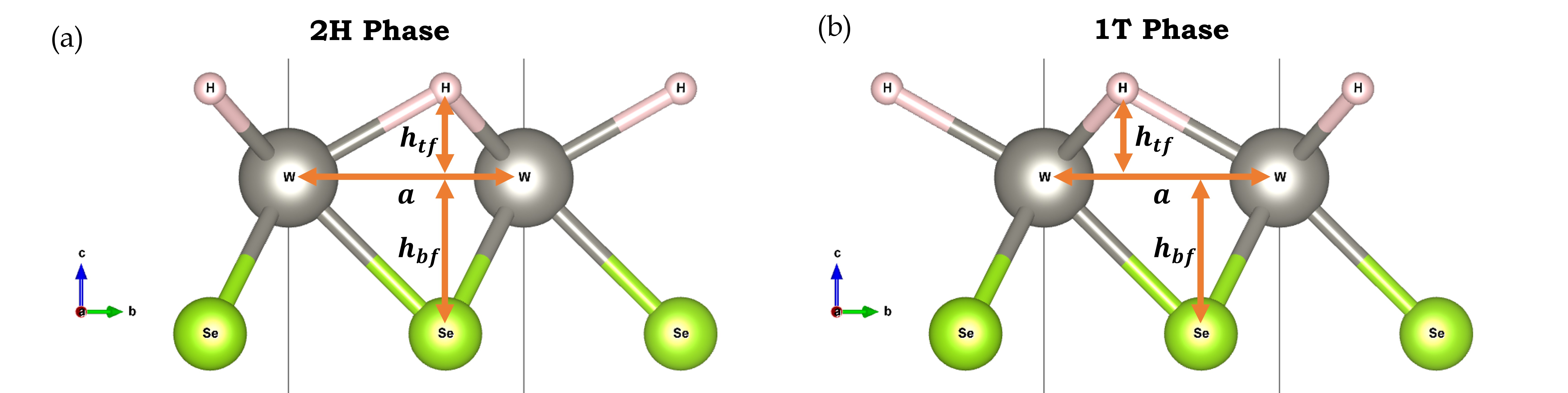}
		\caption{Visualization of the phases of Janus WSeH. (a) The Janus 2H-WSeH consists of the ABA stacking sub-layers of hydrogen (pink), tungsten (grey) and selenium (green). The geometry is determined by the in-plane lattice constant $a$, and the heights of the top and bottom sub-layers with respect to the W sub-layer, which are denoted by $h_{tf}$ for the H sub-layer and $h_{bf}$ for the Se sub-layer. (b) The 1T structure has a similar lattice, but with the ABC stacking.}
		\label{2h-1t-wseh-sideview}
	\end{figure}
 
    In this section, we investigate the structures of the 2H and 1T phases of WSeH and WSH, and compared them with 2H-WSH from a previous work \cite{seeyangnok2024superconductivity}. In addition, we investigated the effects of the applied in-plane strain on the crystal structures. In this work, we concentrated on only one layer of WSeH or WSH, called a monolayer. However, in a WSeH or WSH monolayer, it can be divided into three sub-layers, which are the tungsten, hydrogen, and chalcogenide sub-layers. We chose the tungsten sublayer as the plane of reference. We chose the hydrogen sublayer as the top sublayer and the chalcogenide sublayer as the bottom sublayer. The crystal structures of WSeH are shown in Fig.~\ref{2h-1t-wseh-sideview} as an example. With periodic boundary conditions, the structure belongs to the 3D trigonal space group P3m1 (No.156). For the 2H phase, the Wyckoff positions of tungsten are at (0,0) in the in-plane 2D coordinates, and hydrogen and selenium are at the same Wyckoff positions at (1/3,2/3) in the 2D coordinates. This is called ABA stacking. For the 1T phase, the Wyckoff position of hydrogen changes to (2/3,1/3), and the stacking order becomes ABC.  
    
    For the non-strained study, the optimization of Janus 2H-WSeH gives the lattice constant of $a$= 3.070 \AA , the distance between the top sublayer and the tungsten sublayer is $h_{tf}$= 1.008 \AA , and the bottom sublayer and the tungsten sublayer is $h_{bf}$= 1.764 \AA. When spin-orbit coupling (SOC) is activated, we obtain slightly different values of the cell parameters, which are $a$= 3.070, $h_{tf}$= 1.012 \AA, and $h_{bf}$= 1.769 \AA. For the biaxial strain study, we analyzed various amounts of strain for WSeH and WSH. We found that $h_{tf}$ and $h_{bf}$ decrease as the tensile (positive) biaxial strain is applied, but $h_{tf}$ and $h_{bf}$ increase as the compressive (negative) biaxial strain is applied. All crystal parameters of the non-strain and biaxial strain of the 1T and 2H phases are also summarized in Table \ref{lattice-consts-table}.
 
\begin{table}[b]
\caption{\label{lattice-consts-table}%
The table shows the values of biaxial strain $\epsilon$ (\%), the lattice constant $a$, the distance between the top sub-layer  and the tungsten sub-layer, $h_{tf}$, and the bottom sub-layer and the tungsten sub-layer, $h_{bf}$, of the 1T and 2H phases of WSeH and WSH. The values for the spin-orbit coupling (SOC) case are also reported.
}
\begin{ruledtabular}
\begin{tabular}{ccccc}
\textrm{Compounds}&
\textrm{$\epsilon$ (\%)}&
\textrm{$a$ (\AA)}&
\textrm{$h_{tf}$ (\AA)}&
\textrm{$h_{bf}$ (\AA)}\\
\colrule
2H-WSeH(SOC) & 0 & 3.070 & 1.012 &1.769 \\
2H-WSeH(nonSOC) & -1 & 3.040 & 1.028 & 1.778 \\ 
	& 0 & 3.070 & 1.008 &1.764 \\
	& 1 & 3.101 & 0.988 & 1.751 \\
 \hline
 2H-WSH(SOC) & 0 & 2.994 & 1.069 & 1.634 \\
2H-WSH(nonSOC)  & -1 & 2.963 & 1.085 & 1.644 \\
	& 0 & 2.993 & 1.065 & 1.632 \\
	& 1 & 3.023 & 1.047 & 1.621 \\
	& 2 & 3.052 & 1.030 & 1.610 \\
	& 3 & 3.083 & 1.014 & 1.599 \\
	& 4 & 3.113 & 1.000 & 1.589 \\
	& 5 & 3.143 & 0.987 & 1.579 \\
\hline
1T-WSeH(SOC) & -4 & 2.954 & 1.059 & 1.832 \\
        & -5 & 2.923 & 1.079 & 1.851 \\
1T-WSeH(nonSOC) & 0 & 3.076 & 0.959 & 1.759 \\
        & -4 & 2.953 & 1.058 & 1.829 \\
        & -5 & 2.922 & 1.077 & 1.848 \\
        & -6 & 2.892 & 1.095 & 1.866 \\
\hline
1T-WSH(SOC)	& -2 & 2.939 & 1.076 & 1.655 \\
	& -3 & 2.909 & 1.100 & 1.669 \\
1T-WSH(nonSOC)  & 0 & 2.997 & 1.025 & 1.627 \\
	& -1 & 2.967 & 1.051 & 1.641 \\
	& -2 & 2.937 & 1.075 & 1.654 \\
	& -3 & 2.907 & 1.100 & 1.667 \\
	& -4 & 2.877 & 1.122 & 1.681 \\
	& -5 & 2.847 & 1.144 & 1.695 \\
\end{tabular}
\end{ruledtabular}
\end{table}

	\subsection{Electronic structures}
	
\begin{figure}[h!]
    \centering
		\includegraphics[width=8cm]{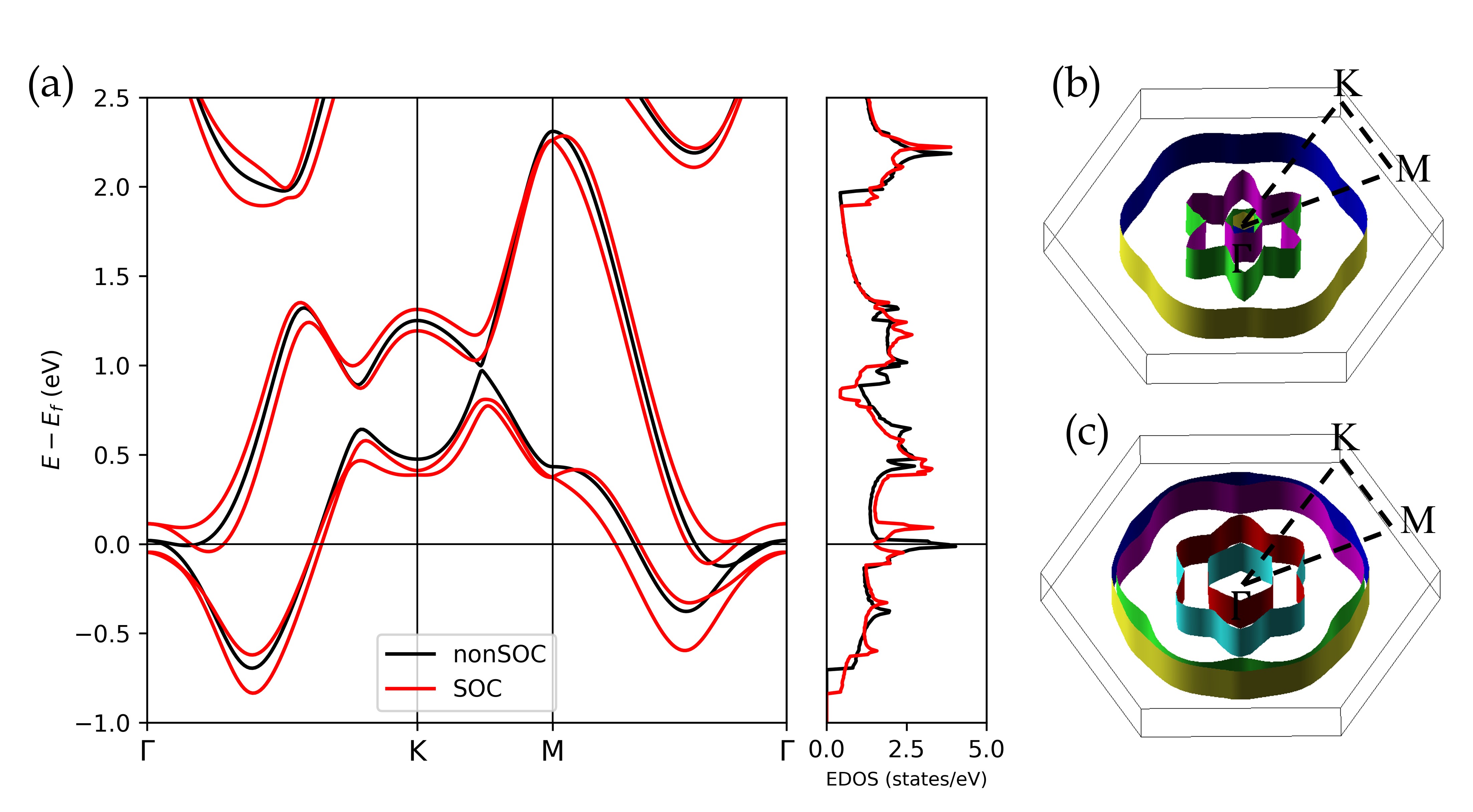}
		\caption{shows the electronic structures, the total density of states and the Fermi surfaces of 2H-WSeH. (a) shows the dispersion of the non-SOC (black lines) and SOC (red lines) electronic structures. The corresponding total DOS are also shown. (b) and (c) show the visualization of the Fermi surfaces of the non-SOC and SOC 2H-WSeH, respectively.}
		\label{2h-estructure-(non)soc}
	\end{figure}

	In this section, we report the results of the electronic structures of WSeH and WSH. There are a few combinations of different lattice structures, such as 2H and 1T structures, and the different electronic structures, such as spin orbit coupling (SOC) and non-SOC electronic structures. The electronic dispersions and the corresponding total electronic density of states (TDOS) from 2H-WSeH with non-SOC (black lines) and SOC (red lines) calculations are shown in Fig.\ref{2h-estructure-(non)soc} (a). It is readily seen that the 2H-WSeH monolayer exhibits metallic behavior as a result. The bands crossing the Fermi level are the W-$d$ orbitals: they have only a small projection onto  the H-$s$ and Se-$p$ orbitals, The d-bands can be grouped into $A'(d_{z^2})$, $E'(d_{xy}, d_{x^2-y^2})$ and $E''(d_{yz}, d_{xz})$ at the $\Gamma$ point. The corresponding Fermi surfaces are shown in Fig.\ref{2h-estructure-(non)soc} (b) and (c) for the case of non-SOC and SOC calculations, respectively. In general, the electronic structures near the Fermi level are very sensitive to the spin-orbit coupling (SOC) effect. In the non-SOC calculation (black lines in Fig.\ref{2h-estructure-(non)soc} (a)), the band dispersions become flat near the $\Gamma$ point, resulting in a sharp peak of TDOS close to the Fermi level, which could be identified as the van Hove singularity (vHs). When we activate the SOC effect, the band degeneracies are removed, especially near the $\Gamma$ point and the Fermi level. Consequently, the vHs peak splits into two smaller vHs peaks near the Fermi level, as shown by the red lines in Fig.\ref{2h-estructure-(non)soc} (a). The SOC effect significantly changes the topology of the Fermi surfaces, as shown in Fig.\ref{2h-estructure-(non)soc} (c) compared to the non-SOC Fermi surfaces in (b). The topology of the Fermi surface is one of the key ingredients in superconductivity. For a conventional superconductor with strong electron-phonon interaction, a high electronic density of states at the Fermi level and strong Fermi surface nesting are the most expected features from the electronic part. The Fermi surface nesting describes the connection between any two different points on the Fermi surfaces. In these compounds, the Fermi surfaces are in the form of several enclosed loops. These loops have some certain spacing between them in the k-space. These features will also manifest themselves in the features of the phonon linewidths. We will discuss these features again when we consider the phonon linewidths in the next sections.
   
   \begin{figure}[h!]
		\centering
		\includegraphics[width=8cm]{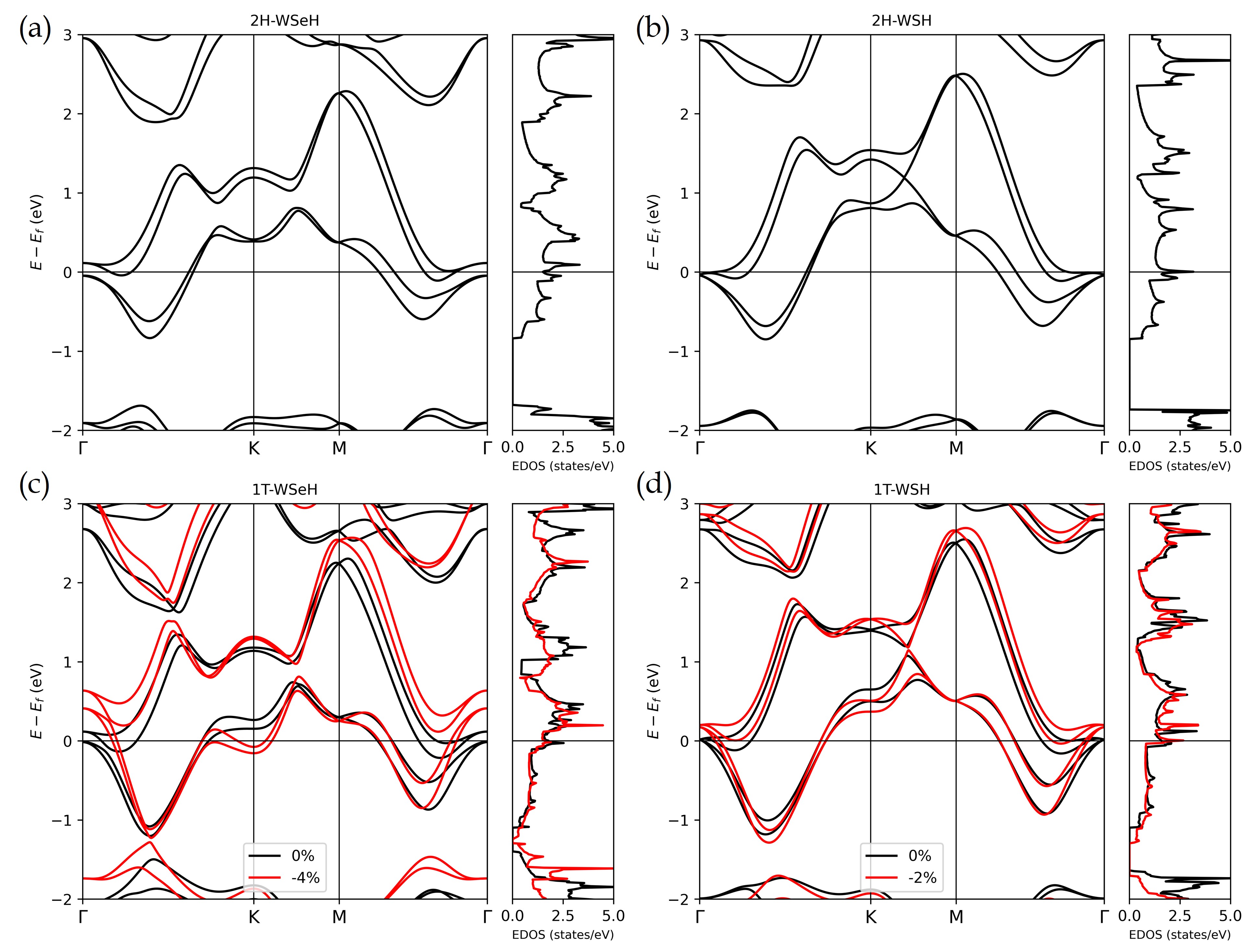}
		\caption{shows the electronic structures and the total density of states of WSeH and WSH with SOC. (a) shows the electronic structure of 2H-WSeH. (b) 1T-WSeH with 0\% (black lines) and -4\% (red lines) biaxial compressive strain. (c) shows the electronic structure of 2H-WSH. (d) 1T-WSH with 0\% (black lines) and -2\% (red lines) biaxial compressive strain.}
		\label{ebands-wxh-1t-2h}
    \end{figure}    
	
    As the SOC effect is the most accurate representation of the electronic structures, we performed all of our calculations with SOC from now on, unless otherwise stated. Fig.\ref{ebands-wxh-1t-2h} (a)-(d) shows the comparison of the electronic structures between the 1T and 2H phases of WSeH and WSH. In general, the electronic structures show metallic behavior, and the bands are dominated by the W-$d$ orbitals that cross the Fermi level. The electronic structures between the 2H and 1T phases are very similar. The dispersions of 1T-WSeH (Fig.\ref{ebands-wxh-1t-2h} (c)) and 1T-WSH (Fig.\ref{ebands-wxh-1t-2h} (d)) below the Fermi level show greater dips to lower energy along $\Gamma$ to $K$, and $\Gamma$ to $M$, compared to 2H-WSeH (Fig.\ref{ebands-wxh-1t-2h} (a)) and 2H-WSH (Fig.\ref{ebands-wxh-1t-2h} (b)). These dips indicate stronger bonding between the H atom at the 1T site and the W atom. These findings should explain why the 1T structures have slightly lower energy than the 2H structures; see Fig.~\ref{fig:energybarrier}. However, when we performed the phonon calculations, we found that the 2H phonons are all stable, but some of the 1T phonons are unstable. Therefore, we applied the biaxial compressive strain to the 1T structure. We found that a small amount of strain can stabilize the 1T phase in both WSeH and WSH, shifting the bands at $\Gamma$ away from the Fermi surface as shown in Fig.\ref{ebands-wxh-1t-2h}. The dynamical stability will be discussed in more detail in the next section. An interesting feature is observed along the $K-M$ directions at +1eV: the bands cross in the Se compounds, but not in the S compounds. These bands have $d$-orbital character, so in WSeH they hybridize between W and Se, breaking symmetry and opening a pseudogap. By contrast, in sulfur has no $d$-electrons, so the symmetry is unbroken and the bands cross.
	
	\section{Phase stability, phonons, and superconductivity}
        In this section, we discuss the phase stability of WSeH and WSH. Then, we calculated the phonons and their associated linewidths from the electron-phonon interaction, and reported the electron-phonon coupling in each phonon bands as a function of wave vectors. Finally, we evaluated the superconducting transition temperatures of these compounds.
        \subsection{Phase stability}
        \begin{figure}[h!]
		\centering
		\includegraphics[width=8.5cm]{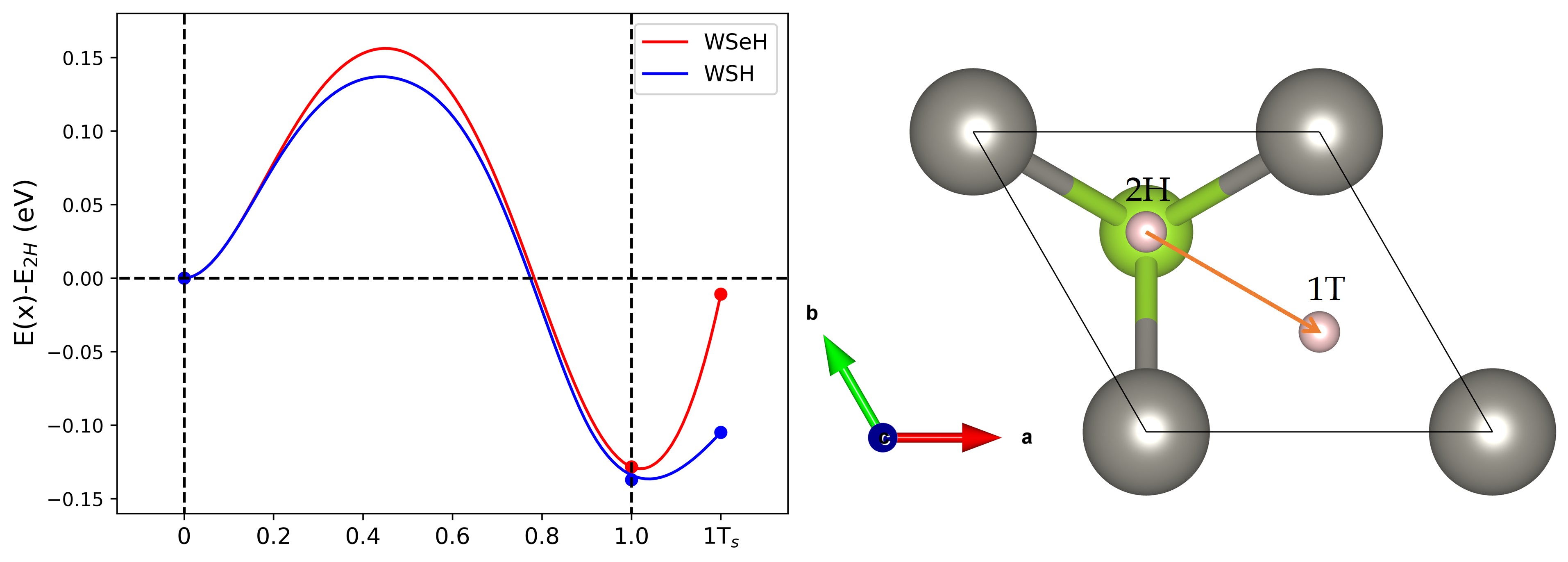}
		\caption{shows the relative energies ($E(x) - E_{2H}$) per formula unit along the parameter $x$, describing the displacement of all hydrogens along the fractional coordinates of the primitive vectors $\vec{a}$ and $\vec{b}$, given by the position vector $\vec{x}=(\frac{(x+1)}{3}\vec{a}, \frac{(2-x)}{3}\vec{b})$, starting from the optimized 2H phase of WSeH (red line) and WSH (blue line) by fixing the z position. The dots denote the optimized structure energies of the 2H and 1T phases, where strain-stabilized 1T structures are denoted by $1T_{s}$ for WSeH ($\epsilon = -4\%$) and WSH ($\epsilon = -2\%$).}
		\label{fig:energybarrier}
	\end{figure}
    
    At this stage, we discuss the static phase stability of WSeH and WSH. In Fig.\ref{fig:energybarrier}, we found that the 1T-WSeH and 1T-WSH monolayers are more energetically favorable than their 2H phases. For the WSeH monolayer, the energy of the 1T phase is lower than that of the 2H phase by 0.1283 eV. For the WSH monolayer, the energy of the 1T phase is lower than that of the 2H phases by 0.1369 eV. However, we also found that the 1T-WSeH and 1T-WSH structures are not dynamically stable unless compressive biaxial strain is applied. The strain-stabilized 1T structure will be referred to as $1T_{s}$. We found that $1T_{s}$-WSeH with compressive biaxial strain of 4\% has a 0.0109 eV lower energy than that of 2H-WSeH, and $1T_{s}$-WSH with compressive biaxial strain of 2\% is 0.1048 eV has a lower energy than that of 2H-WSH. The energy barriers for WSeH and WSH for the phase transition from 2H to 1T and a strain-stabilized 1T are shown in Fig.~\ref{fig:energybarrier}. It is clearly seen that relatively small amounts of energy can drive the phase transformation from the 2H to the 1T phase. It is also worth noting that both the 2H and 1T structures may not be true ground-state structures. However, the search for the ground-state structure is beyond the scope of this work. Instead, we tried to show that both phases could possibly be synthesized as in the case of MoSH \cite{lu2017janus}. Hence, we determined the formation energy of the crystal structure given by \begin{equation}            
        E_{\text{formation}}=E_{\text{structure}}-E_{\text{W}}-E_{\text{Se/S}}-E_{\text{H}},
    \end{equation}
    where $E_{\text{W}}$, $E_{\text{Se/S}}$ and $E_{\text{H}}$ are the energy of the isolated tungsten, chalcogenide (Se/S) and hydrogen atoms, respectively. The formation energies per formula unit of 2H-WSeH and 2H-WSH are -15.86 and -17.24eV, while the formation energies of 1T-WSeH with strain -4\% and 1T-WSH with strain -2\% are -15.87 and -17.35eV, respectively. 
        
	\subsection{Phonons}
	\begin{figure}[h!]
		\centering
		\includegraphics[width=8cm]{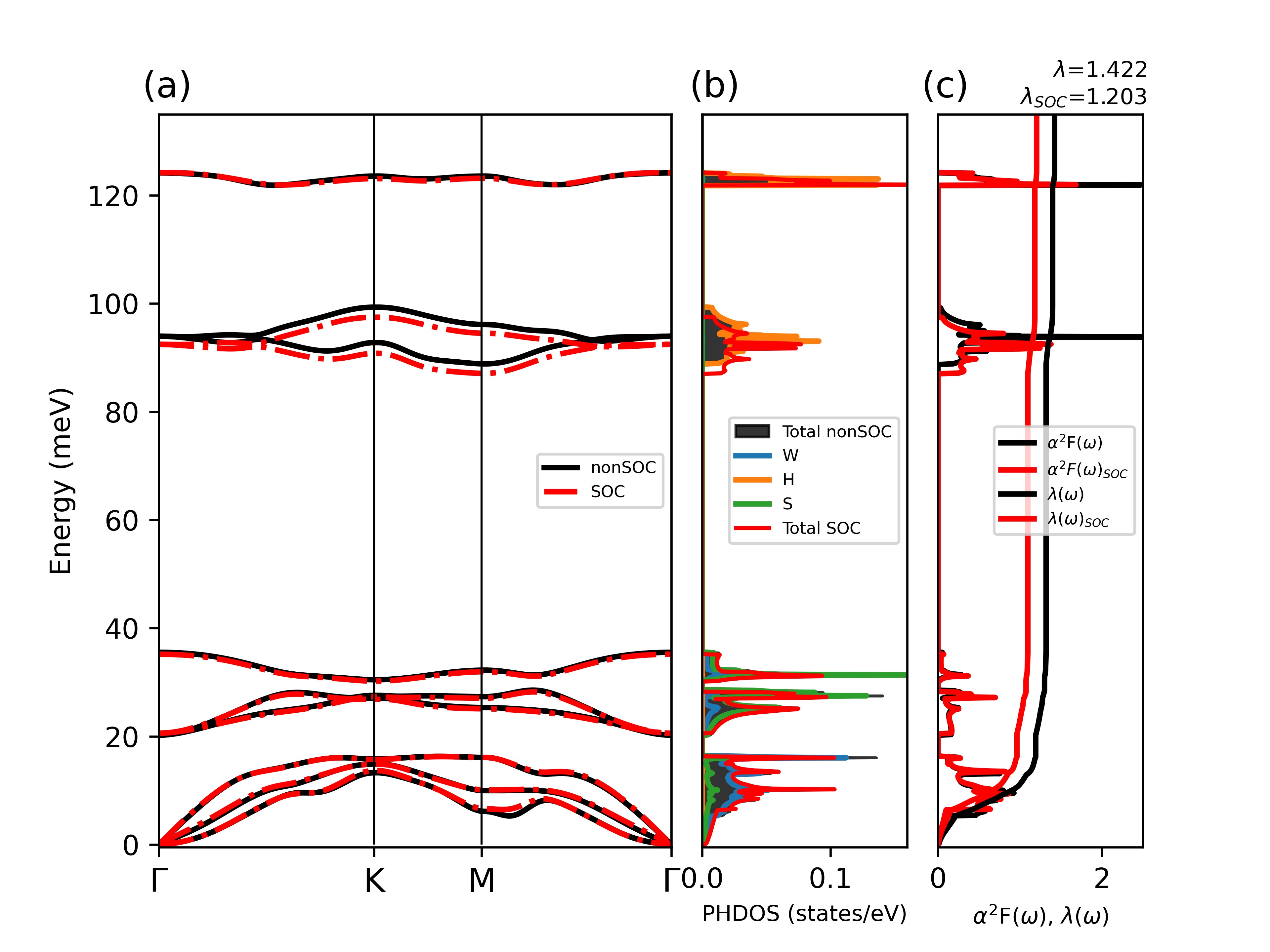}
		\caption{shows (a) the phonon dispersions of the non-SOC (black lines) and SOC (dashed red lines) 2H-WSeH, (b) the phonon density of states of the non-SOC (black lines) and SOC (red lines)  2H-WSeH, and the projected phonon density of states of W (blue lines), H (orange lines), and S (green lines) atoms, (c) the isotropic Eliashberg spectral function and the cumulative electron-phonon coupling strength of the non-SOC (black lines) and SOC (red lines) 2H-WSeH.}
		\label{2h-wseh-ph-elph}
	\end{figure}

     Firstly, we discussed the phonons of the 2H WSeH. The calculations were performed using both non-SOC and SOC schemes. Fig.\ref{2h-wseh-ph-elph}(a) shows that all phonon modes are stable for this compound with the P3m1 trigonal space group. As the primitive cell contains three atoms, the phonon bands consist of three acoustic branches ($\nu = 1-3$) and six optical branches ($\nu = 4-9$). The three acoustic branches can be divided into an in-plane longitudinal mode (LA), an in-plane transverse mode (TA), and an out-of-plane flexural mode (ZA). These acoustic phonon bands will be referred to as band $\nu = 1-3$. These bands occupy the frequency range of 0-15meV. The dispersion of the LA and TA modes must have linear behavior at the long-wavelength limit, while the dispersion of the ZA must exhibit a quadratic relation, as expected in all 2D materials \cite{molina2011phonons,zhu2014coexistence,zhang2014thermal,taheri2021importance,lin2022general}. For the six optical modes, they can be categorized into two sets of one $A_{1}$ and two $E$ subgroups. The first set of the optical bands will be referred to as band $\nu = 4-6$. These bands occupy the frequency range of 20-35meV. The second set of the optical bands will be referred to as band $\nu = 7-9$. The band 7 and 8 occupy the frequency range of 85-100meV, while the band 9 occupies the frequency range of around 125 meV. All bands and their associated frequency ranges are shown in Fig.\ref{2h-wseh-ph-elph} (b). The absence of the 3D inversion symmetry causes all modes to be Raman and IR active. All important information is summarized in Table \ref{tab:phonon-eigenvalues} for the SOC scheme. The most significant effects of the SOC scheme on the phonons are that the frequencies of the ZA modes around the $M$ point are a little higher than those of the non-SOC scheme, and the frequencies of the $E$ optical modes (the whole band 7 and 8) are a little lower than those of the non-SOC scheme. These behaviors are similar to those of 2H-WSH \cite{seeyangnok2024superconductivity}. The noticeable differences between the phonons of WSeH and WSH are the frequencies of the band $\nu = 4-6$, where the eigenvectors are dominated by the vibrations of the S/Se atoms. As the S atom is 2.46 times lighter than the Se atom, the frequencies of these modes (band $\nu = 4-6$) in WSH are $\sqrt{2.46}=1.56$ times higher than those of WSeH; see Table~\ref{tab:phonon-eigenvalues} for comparison. We also studied the effects of strain on the 2H phase. We found that the 2H phase can withstand small amounts of both tensile and compressive strain, maintaining positive phonon frequencies. The corresponding superconducting quantities, computed under these strains, are shown in Table~\ref{tab:2h-strain-enhanced-sc}.

\begin{table}[b]
\caption{\label{tab:phonon-eigenvalues}
summarizes all the important information of the phonon modes, such as band numbers, subgroups, eigenvectors, infrared (I) or Raman (R) active, the frequencies (meV) at the $\Gamma$ point of both 2H and 1T structures. These values are calculated in the SOC scheme. The upper table reports the values of 2H and -4\% strain-stabilized 1T-WSeH, whereas the lower table reports the values of 2H and -2\% strain-stabilized 1T-WSH.
}
\begin{ruledtabular}
\begin{tabular}{cccccccc}
\textrm{Band $\nu$}&
\textrm{Subgroup}&
\textrm{Eigenvector}&
\textrm{Active}&
\textrm{2H}&
\textrm{1T}\\
\colrule
		4,5 & E & \text{In-plane Se, H} & I+R & 20.57 & 22.31  \\
		6 & $A_{1}$ & \text{Out-plane Se, H} & I+R & 35.18 & 33.03  \\
		7,8 & E & \text{In-plane H} & I+R & 92.24 & 113.70  \\
		9 & $A_{1}$ & \text{Out-plane H}  & I+R & 124.23 & 129.07  \\
\colrule

  	4,5 & E & \text{In-plane S, H} & I+R & 31.50 & 32.87  \\
		6 & $A_{1}$ & \text{Out-plane S, H} & I+R & 54.36 & 54.57  \\
		7,8 & E & \text{In-plane H} & I+R & 94.78 & 116.25  \\
		9 & $A_{1}$ & \text{Out-plane H}  & I+R & 130.91 & 130.94  \\
\end{tabular}
\end{ruledtabular}
\end{table}    

	\begin{figure}[h!]
		\centering
		\includegraphics[width=8cm]{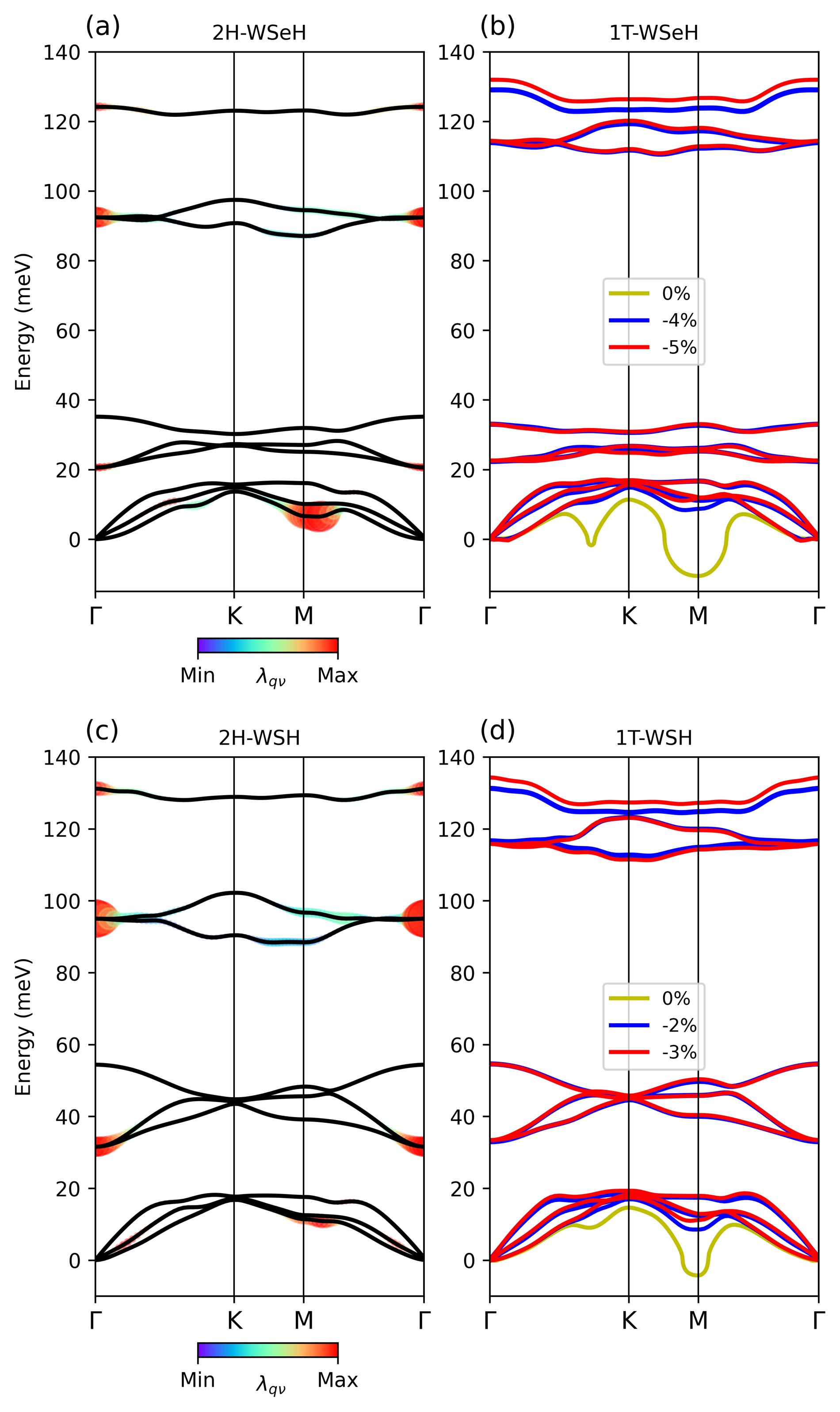}
	\caption{shows (a) the phonon dispersion of 2H-WSeH, accompanied by painted and broadened areas 
 showing where the electron-phonon coupling $\lambda_{\bold{q}\nu}$ are most significant.  To emphasize the most important areas, 
 the size of the painted region is proportional to $\lambda_{\bold{q}\nu}^{4}$.
 (b) the phonon dispersions of unstrained (0\%, green lines), -4\% (blue lines), and -5\% (red lines) strained 1T-WSeH, (c) the the weighted electron-phonon coupling phonon dispersion of 2H-WSH,  accompanied by the painted areas in which $\lambda_{\bold{q}\nu}$ are most significant, (d) the phonon dispersions of unstrained (0\%, green lines), -2\% (blue lines) and 3\% (red lines) strained 1T-WSH. All calculations were performed with the SOC scheme. It is worth noticing that the unstrained 1T structures are dynamically unstable.}
		\label{phband-2h1t}
	\end{figure}

\begin{figure}[h!]
		\centering
		\includegraphics[width=8cm]{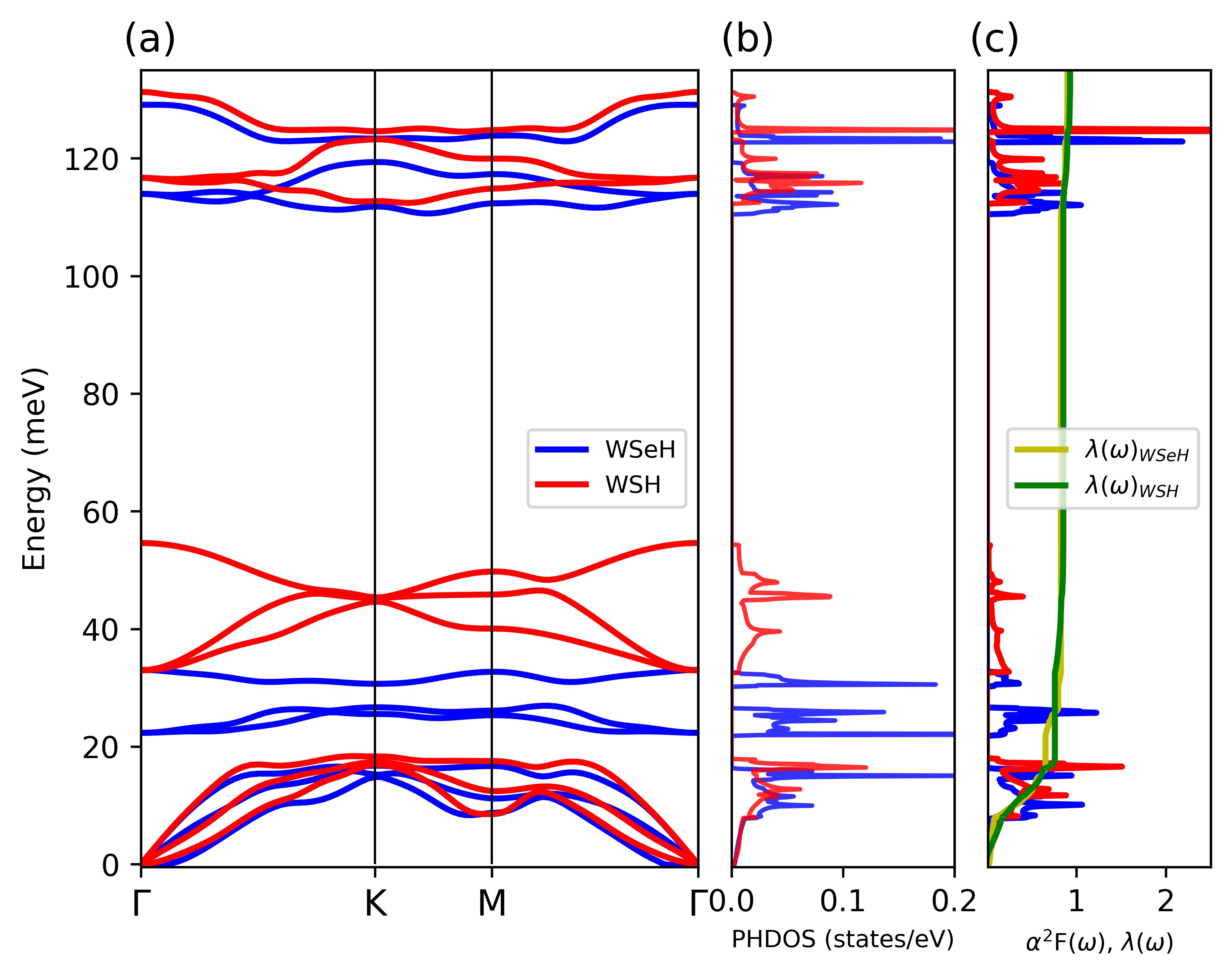}
		\caption{shows (a) the phonon dispersions, (b) the phonon density of states, and (c) the isotropic Eliashberg spectral functions of the 1T-WSeH with -4\% strain (blue lines) and 1T-WSH with -2\% strain (red lines). The cumulative electron-phonon coupling strengths are also shown for 1T-WSeH (light green line) and 1T-WSH (green line). All calculations were performed with the SOC scheme.}
		\label{1t-phbands-elph}
	\end{figure}

    Next, we investigate the 1T structures. From Fig.\ref{phband-2h1t} (b) and (d), it is obvious that the unstrained 1T structures (green lines) are dynamically unstable in both WSeH and WSH. Thus, this leads us to consider the strained 1T structures. In this work, we consider only the biaxial strain. For the 1T structures, we found that -4\% biaxial strain (blue lines) is needed to stabilize the 1T-WSeH structure, and -2\% biaxial strain (blue lines) is needed to stabilize the 1T-WSH structure, as shown in Fig.\ref{phband-2h1t} (b) and (d). If we try to increase compression, the frequencies of the phonons tend to increase slightly, as shown by the red lines in Fig.\ref{phband-2h1t} (b) and (d). The main differences between the 1T and 2H structures are the frequencies of the phonon band 7 and 8. From the eigenvectors, the vibrations of these bands are dominated by the movement of the H atom, and somehow the 1T structures give higher frequencies of these modes than their 2H structures; see Fig.\ref{phband-2h1t} (a) and (b), and Fig.\ref{phband-2h1t} (c) and (d), and also Table~\ref{tab:phonon-eigenvalues} for comparison. These higher frequencies should come from the stronger bonding between the H atom at the 1T site and the W atom, as discussed in the section on electronic structures. The stable phonons of the 1T phase of -4\% strained WSeH (blue lines) and -2\% strained WSH (red lines) are compared again in Fig.~\ref{1t-phbands-elph} for clarification. The electron-phonon coupling ($\lambda_{\bold{q}\nu}$) is also projected onto the phonon dispersion in the 2H phase of WSeH and WSH, as shown in Fig.\ref{phband-2h1t} (a) and (c), which will be discussed in the next section along with Fig.~\ref{lambda-map2D}. It shows that one of the most strongly coupled phonons arises from the $M$ point, which will play a key role in superconductivity as the electron-phonon coupling ($\lambda_{\bold{q}\nu}$) is averaged to $\lambda$.

 \begin{figure}[h!]
		\centering
		\includegraphics[width=8.5cm]{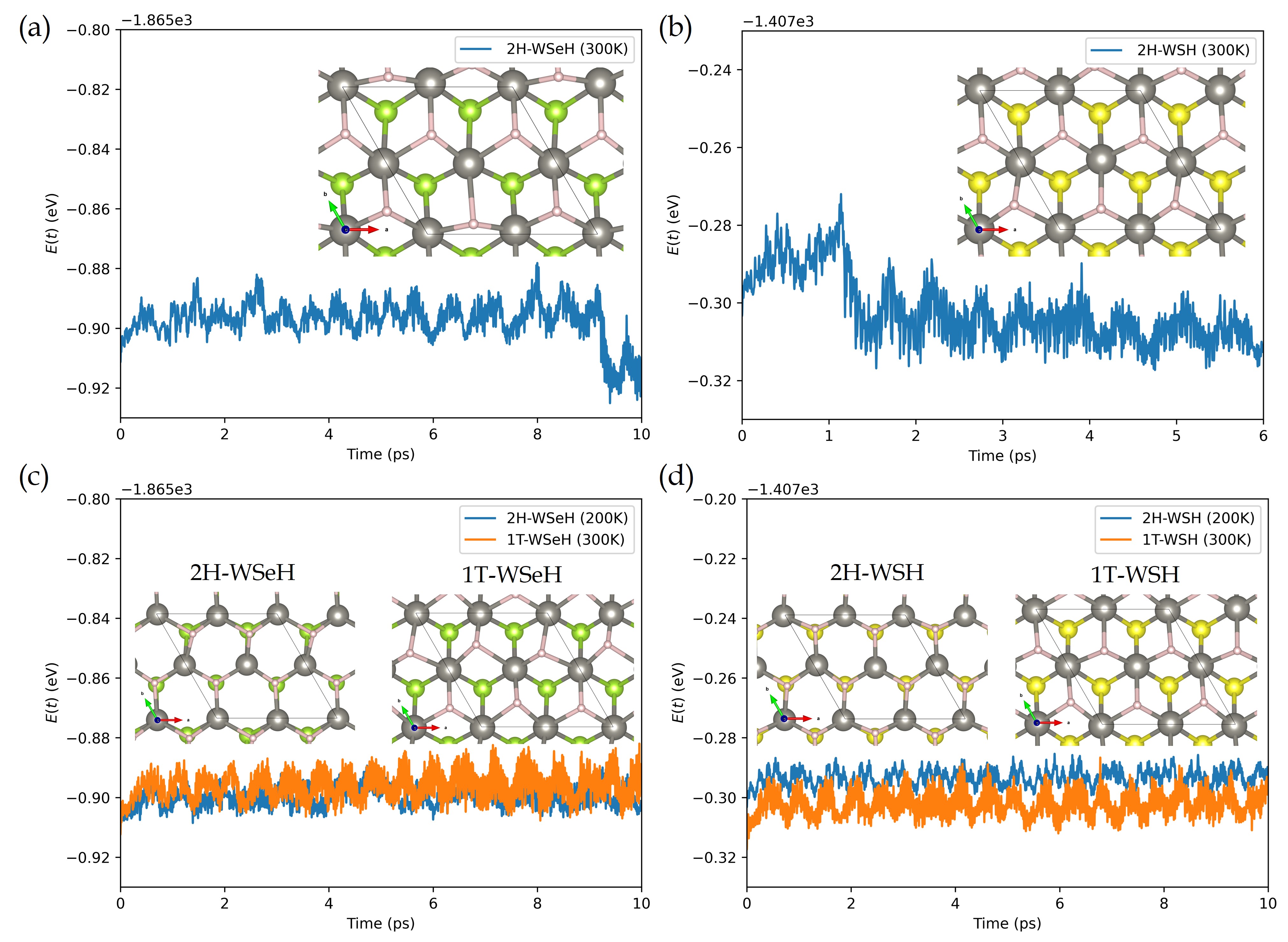}
		\caption{shows the AIMD results for a $2\times2\times1$ 2H/1T WSH supercell using Born-Oppenheimer molecular dynamics (BOMD) implemented in CASTEP \cite{CASTEP}. Figures (a) and (b) illustrate the time evolution in the NVT ensemble for the initial 2H phase structures of WSeH and WSH at 300 K, respectively. The energy dips around 1 ps and 9 ps indicate a phase transition from 2H to 1T. Figures (c) and (d) depict the NVT ensemble results for initial 2H (blue line) and 1T (orange line) phases at 200 K and 300 K, respectively, for both WSeH and WSH. The snapshot configurations on the left and right were captured at the final timesteps of the 2H and 1T initial structures, respectively, demonstrating that both phases remain unchanged after 10.0 ps.}
  	\label{fig:MD-plos}
	\end{figure}

    As the unstrained 1T has unstable phonons, we investigated anharmonicity using  \textit{ab initio} molecular dynamics (AIMD) simulations. This is to determine whether  thermal effects can stabilize the high-symmetry phase.  We use the Born-Oppenheimer approximation which means that the zero-point motion is ignored. Zero-point motion can stabilize phonons which are unstable ("imaginary") in lattice dynamics \cite{drummond2002ab}.  It also means that tunnelling through barriers is not possible. Here, we will demonstrate that the imaginary phonons are dynamically stabilised by temperature: this gives an upper bound since zero point would only make them more stable. BOMD may overestimate the metastability of the 2H phase, but the barrier calculation (Fig.~\ref{fig:energybarrier}) suggest that the tunnelling  barrier is both high and wide.
    
    The results of the AIMD simulations in the NVT ensemble are shown in Fig.~\ref{fig:MD-plos}.
    The 2H phase of both WSH and WSeH remain stable at 200K for 10ps, while for the 1T phase, both WSH and WSeH, with applied strains of -2\% and -4\%, respectively, shows thermal stability at room temperature of 300K where the result of WSH is shown in Fig.~\ref{fig:MD-plos} (c) and (d). 
    
    At  300K, for both WSH and WSeH, we observed that the hydrogen atoms begin to migrate from the 2H sites to the 1T sites. This indicates that the 2H phase starts to undergo a phase transformation to the 1T phase, see Fig.~\ref{fig:MD-plos} (a) and (b). For the 1T phase, we found that with a suitable lattice constant, the 1T phase of WSeH and WSH is stable up to room temperature. This finding is consistent with the calculations of the static structures, which has 1T as the lower energy phase and a barrier of around 150meV as shown in Fig.~\ref{fig:energybarrier}.

\subsection{Phonon linewidths and electron-phonon coupling}
    
    \begin{figure}[h!]
		\centering
		\includegraphics[width=8cm]{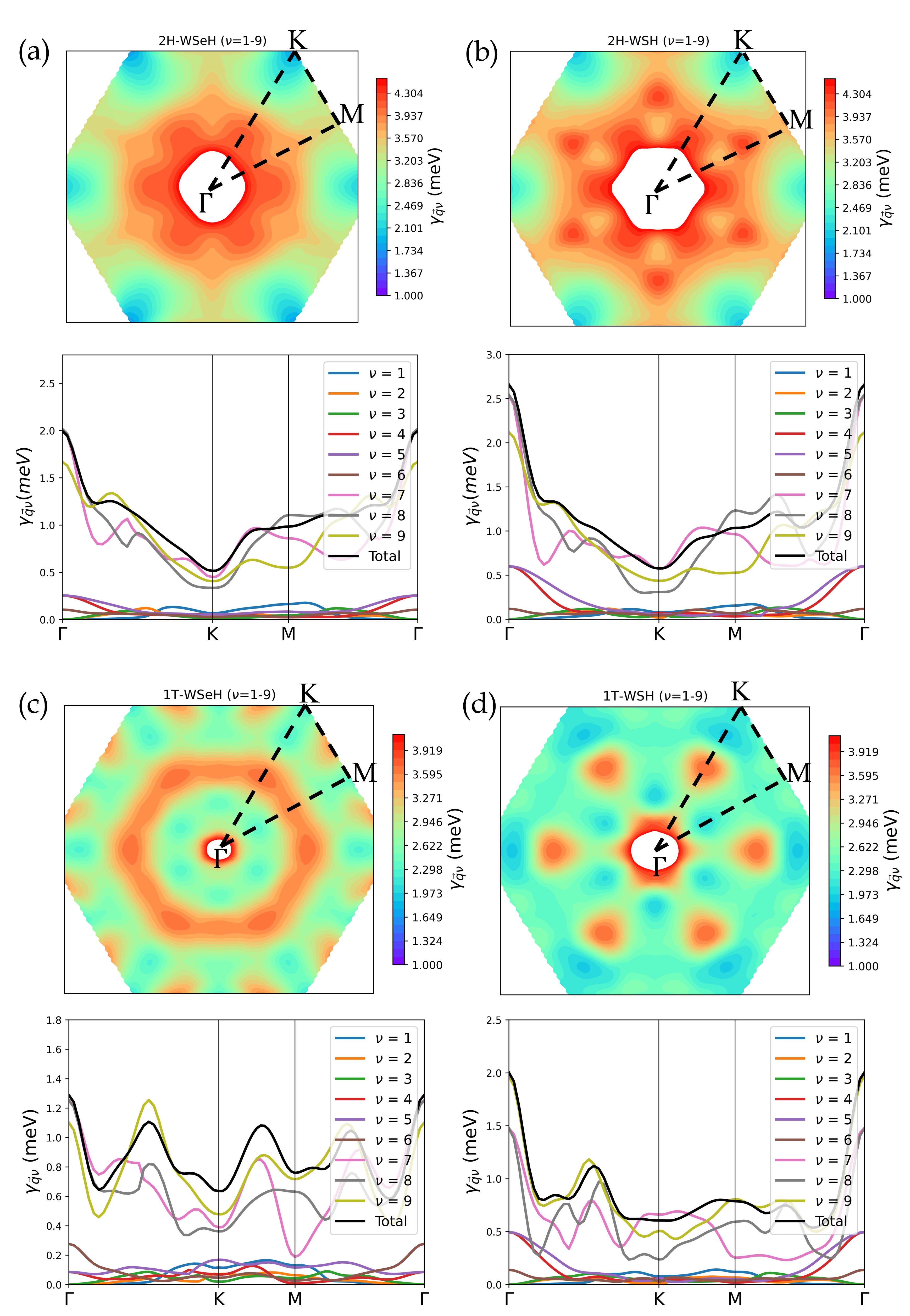}
		\caption{shows the phonon linewidths (Eq.\ref{gammaphononlinewidths}), $\gamma_{\bold{q}\nu}$ of (a) 2H-WSeH, (b) 2H-WSH, (c) 1T-WSeH ($\epsilon=-4\%$), and (d) 1T-WSH ($\epsilon=-2\%$). The lower plots of (a), (b), (c) and (d) show the dispersions of $\gamma_{\bold{q}\nu}$, calculated with the SOC scheme. The dispersions of the different phonon bands $\nu$ are denoted by different colors. The total phonon linewidths (black lines) are divided by 4. The upper plots are the associated color mappings of the total $\gamma_{\bold{q}}$ in $\bold{q}$-space.}
		\label{gamma-2h}
	\end{figure}

 \begin{figure}[h!]
		\centering
		\includegraphics[width=8cm]{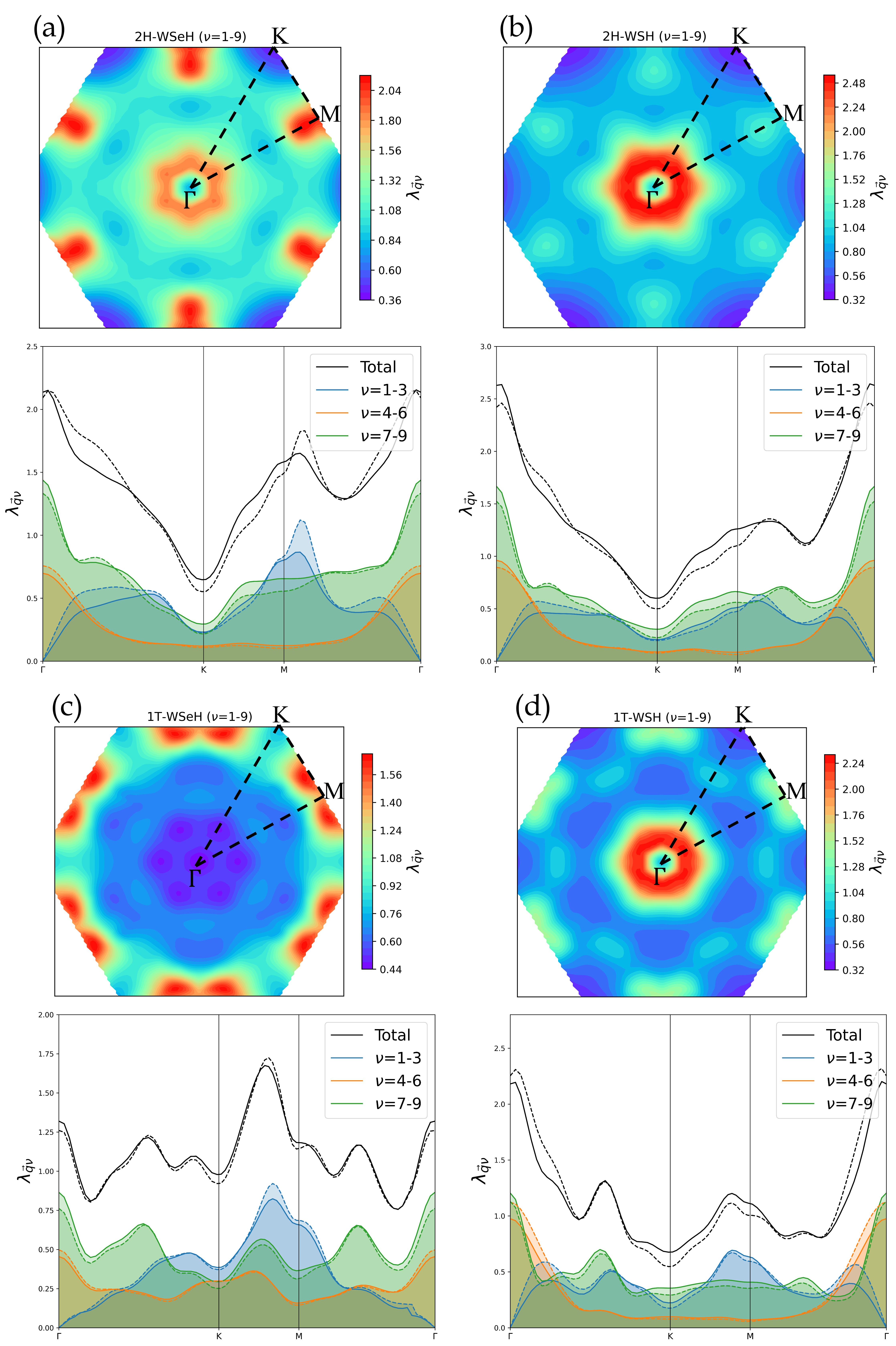}
		\caption{shows the electron-phonon coupling, $\lambda_{\bold{q}\nu}$ (Eq.\ref{eqn:lambda_qv} (a) 2H-WSeH, (b) 2H-WSH, (c) 1T-WSeH ($\epsilon=-4\%$), and (d) 1T-WSH ($\epsilon=-2\%$). The lower plots of (a), (b), (c) and (d) show the dispersions of $\lambda_{\bold{q}\nu}$. The SOC results are shown with full lines, while the non-SOC results are shown with dashed lines. The upper plots are the associated color mappings of the total $\lambda_{\bold{q}}$ in $\bold{q}$-space.}
		\label{lambda-map2D}
    \end{figure}

    In this section, we evaluated the phonon linewidth, $\gamma_{\bold{q}\nu}$, and the electron-phonon coupling. $\lambda_{\bold{q}\nu}$, as a function of phonon wave vector $\bold{q}$ of band $\nu$, where $\nu = 1-9$. From these two equations, it can be derived that $\lambda_{\bold{q}\nu} \propto \frac{\gamma_{\bold{q}\nu}}{\omega^2_{\bold{q}\nu}}$. Due to the electron-phonon interaction, the phonons are treated as a quasiparticle with a finite lifetime. This lifetime is directly related to $\gamma_{\bold{q}\nu}$. From Eq.(\ref{gammaphononlinewidths}), $\gamma_{\bold{q}\nu}$ also gives the description of the weighted connection between any two points on the Fermi surfaces by the phonon wave vector $\bold{q}$ of the band $\nu$. The Fermi surface connection can be written in analytic form as
    \begin{equation}    \delta(\epsilon_{\bold{k}}-\epsilon_{F})\delta(\epsilon_{\bold{k}+\bold{q}}-\epsilon_{F}),
    \end{equation}
    and weighted by the transition propability $|g_{\boldsymbol{k}+\boldsymbol{q},\boldsymbol{k}}^{\boldsymbol{q}\nu,mn}|^{2}$, where $m$ and $n$ are the electronic band index.
    The Fermi surface connection can be divided into two groups; an intraband and an interband connections. For the intraband, $m = n$, the connecting vector $\bold{q}$ can have a magnitude roughly around $0 \lesssim q < 2k_F$. It appears that there are more connections of low q than those of large q, see the peaks near $\Gamma$ point in Fig.\ref{gamma-2h}. All these peaks are from the optical modes, $\nu = 4-9$. The total magnitude of these peaks near $\Gamma$ point is truncated in the color mapping, otherwise it will overwhelm the features of the other peaks in the brillouin zone. However, for the acoustic mode, $\nu = 1-3$, there is a special feature in which $\gamma_{\bold{q}\nu} \rightarrow 0$ as $\bold{q} \rightarrow 0$. For the interband, $m \neq n$, the connecting vector $\bold{q}$ can only have some discrete values, depended on the inter spacing between the different Fermi surfaces. This is because the topology of the Fermi surfaces in these compounds is in the form of several enclosed loops, where there are specific values of spacing between the different surfaces. The connection between these surfaces requires a specific value of $\bold{q}$. Thus, these features will manifest themselves as peaks in $\gamma_{\bold{q}\nu}$ at specific values of $\bold{q}$ only; see examples of the peaks away from the $\Gamma$ point in Fig.\ref{gamma-2h}, or the red spots in the color mappings. 
    
    For $\lambda_{\bold{q}\nu} \propto \frac{\gamma_{\bold{q}\nu}}{\omega^2_{\bold{q}\nu}}$, we can discuss general behaviors in the acoustic and optical modes. As $\lambda = \sum_{\bold{q}\nu} \lambda_{\bold{q}\nu}$, we would expect to find a high value of $\lambda_{\bold{q}\nu}$ everywhere. However, due to specific electronic and phonon structures, these lead to some noticeable patterns in $\lambda_{\bold{q}\nu}$. For the acoustic mode, $\nu = 1-3$,  $\gamma_{\bold{q}\nu} \rightarrow 0$ and $\omega_{\bold{q}\nu} \rightarrow 0$ as $\bold{q} \rightarrow 0$, and it has a small variation elsewhere. If $\gamma_{\bold{q}\nu} \rightarrow 0$ fast enough, $\lambda_{\bold{q}\nu}$ will go to zero as $\bold{q} \rightarrow 0$ as well. Thus, $\lambda_{\bold{q}\nu}$ can be large only if $\omega_{\bold{q}\nu}$ have comparatively low frequencies, as occurred around those softening points in WSeH, see peaks around the M point in Fig.~\ref{lambda-map2D}, both in the dispersions and in the color mappings. For the optical mode, $\nu = 4-9$, $\omega_{\bold{q}\nu}$ usually have high frequencies. Thus, $\lambda_{\bold{q}\nu}$ can be large near the $\Gamma$ point, where $\gamma_{\bold{q}\nu}$ is large, as seen in Fig.~\ref{lambda-map2D}. These mappings of $\lambda_{\bold{q}\nu}$ give the description of how strong the electron-phonon coupling is in the phonon $\bold{q}$-space. 
 
	\subsection{Superconducting temperature}
 \begin{table}[h]
\caption{\label{tab:2h-strain-enhanced-sc}
The table shows the EPC $\lambda$, $\omega_{\text{ln}}$, and T$_{c}$ of the biaxial strained ($\epsilon$) structures of the 2H and 1T phases of WSeH and WSH.
}
\begin{ruledtabular}
\begin{tabular}{ccccc}
\textrm{Compounds}&
\textrm{$\epsilon$ (\%)}&
\textrm{$\lambda$}&
\textrm{$\omega_{\text{ln}}$(meV)}&
\textrm{$T_{c}$(K)}\\
\colrule
    2H-WSeH(SOC)    & 0 & 1.203 & 11.13 & 11.60 \\
    2H-WSeH(nonSOC)         & -1 & 1.351 & 8.893 & 10.59 \\
			    & 0 & 1.422 & 9.224  & 11.58 \\
			    & +1 & 1.754 & 7.683 & 11.64 \\
    \hline
    2H-WSH(SOC)    & 0 & 1.060 & 13.83 & 12.19  \\
    2H-WSH(nonSOC)          & -1 & 1.175 & 11.42 & 11.56 \\
			    & 0 & 1.178 & 11.36  & 11.53 \\
			    & +1 & 1.187 & 12.05 & 12.35 \\
                    & +2 & 1.198 & 12.14 & 12.59 \\
                    & +3 & 1.295 & 11.46 & 13.02 \\
    \hline
    1T-WSeH(SOC) & -4 & 0.899 &  13.56 & 9.23 \\
            & -5 & 0.759 &  16.19 & 7.91 \\  
    1T-WSeH(nonSOC) & -4 & 0.949 &  13.07 & 9.72 \\
            & -5 & 0.790 &  15.33 & 8.13 \\
            & -6 & 0.709 & 16.29  & 6.79 \\  
    \hline
    1T-WSH(SOC)  & -2 & 0.929 & 14.65 & 10.52 \\
            & -3 & 0.759 & 18.61 & 9.06  \\    
    1T-WSH(nonSOC)  & -2 & 1.063 & 11.31 & 10.01 \\
            & -3 & 0.860 & 15.67 & 9.82 \\
            & -4 & 0.719 & 20.56 & 8.85 \\          
\end{tabular}
\end{ruledtabular}
\end{table}
    From the Eliashberg spectral function, $\alpha^{2} F(\omega)$, as shown in Fig.~\ref{2h-wseh-ph-elph} (c) and~\ref{1t-phbands-elph} (c), we can compute the electron-phonon coupling constant $\lambda$ by using Eq.(\ref{computelambdaformulas}) and the logarithmic average of the phonon energy $\omega_{\text{ln}}$ by using Eq.(\ref{computeomegaln}). Then, the superconducting critical temperature can be evaluated by using Eq.(\ref{allencomputetc}). We assumed $\mu^* = 0.1$. All superconducting data are shown in Table~\ref{tab:2h-strain-enhanced-sc}.  There are several notable points, as follows; 
    
    1. The data show that $\lambda$ in the non-SOC scheme is slightly larger than that of the SOC scheme of the same phase of the same compound. 
    
    2. $\omega_{\text{ln}}$ in the non-SOC scheme is slightly lower than that of the SOC scheme of the same phase of the same compound. 
    
    3.  Strain is important in stabilizing the phonons, and $T_c$ is slightly higher in the phase under tensile strain compared to compressive strain.
    But the effect on T$_c$ is small (Table III).
    
    4.  $T_c$ appears to be rather unaffected by spin-orbit coupling. For 2H-WSeH, $T_c$ is 11.58K and 11.60K in the non-SOC and SOC schemes, respectively. In the case of 2H-WSH the values are 12.2K (SOC) and 11.6K (non-SOC)\cite{seeyangnok2024superconductivity}.

	\section{Conclusion}
    In this work, we investigated monolayers of WSeH and WSH. The electronic structures show that WSeH and WSH exhibit metallic behavior with the Fermi level dominated by the d orbitals of tungsten. Spin-orbit coupling (SOC) breaks the degeneracy of the d-bands leading to the topological changes of the Fermi surfaces, but SOC was shown to have small effects on T$_{c}$. The 2H phase has higher energy than the 1T phase. However, the 1T phase itself is dynamically unstable with the imaginary M-point phonons. The stability of the 1T phase can be restored by applying 4\% and 2\% compressive biaxial strain (the 2D equivalent of pressure) for the cases of WSeH and WSH, respectively. MD simulation suggests that anharmonicity also stabilize 1T through thermal vibration.
 We found relatively  energy barriers of approximately 0.16eV and 0.14eV along the path of transformation from the 2H phase to the 1T  for WSeH and WSH, respectively large enough that 2H remains stable in  molecular dynamics (MD) simulations at 200K. In particular, we discovered that the metastable 2H phase of WSeH has T$_{c}$ $\approx$ 11.60K, similar to WSH. The strained 1T structures of WSeH and WSH have T$_{c}$ $\approx$ 9.23K and 10.52K, respectively. Despite some stability issues, the related MoSH has been synthesized under experimental conditions \cite{lu2017janus}. We believe that the strain induced by a substrate would be the key ingredient in the synthesis of this class of 2D materials.

\begin{acknowledgments}
	This research project is supported by the Second Century Fund (C2F), Chulalongkorn University. The authors acknowledge the National Science and Technology Development Agency, National e-Science Infrastructure Consortium (URL:www.e-science.in.th), Chulalongkorn University and the Chulalongkorn Academic Advancement into Its 2nd Century Project (Thailand) for providing computing infrastructure that has contributed to the research results reported within this paper. GJA acknowledges funding from the ERC project Hecate. We also acknowledge the supports from the Cirrus UK National Tier-2 HPC Service at EPCC (http://www.cirrus.ac.uk), funded by the University of Edinburgh and EPSRC (EP/P020267/1).
 
\end{acknowledgments}

\section*{Appendix A: Projected density of states}
\begin{figure}[H]
    \centering
    \includegraphics[width=7.5cm]{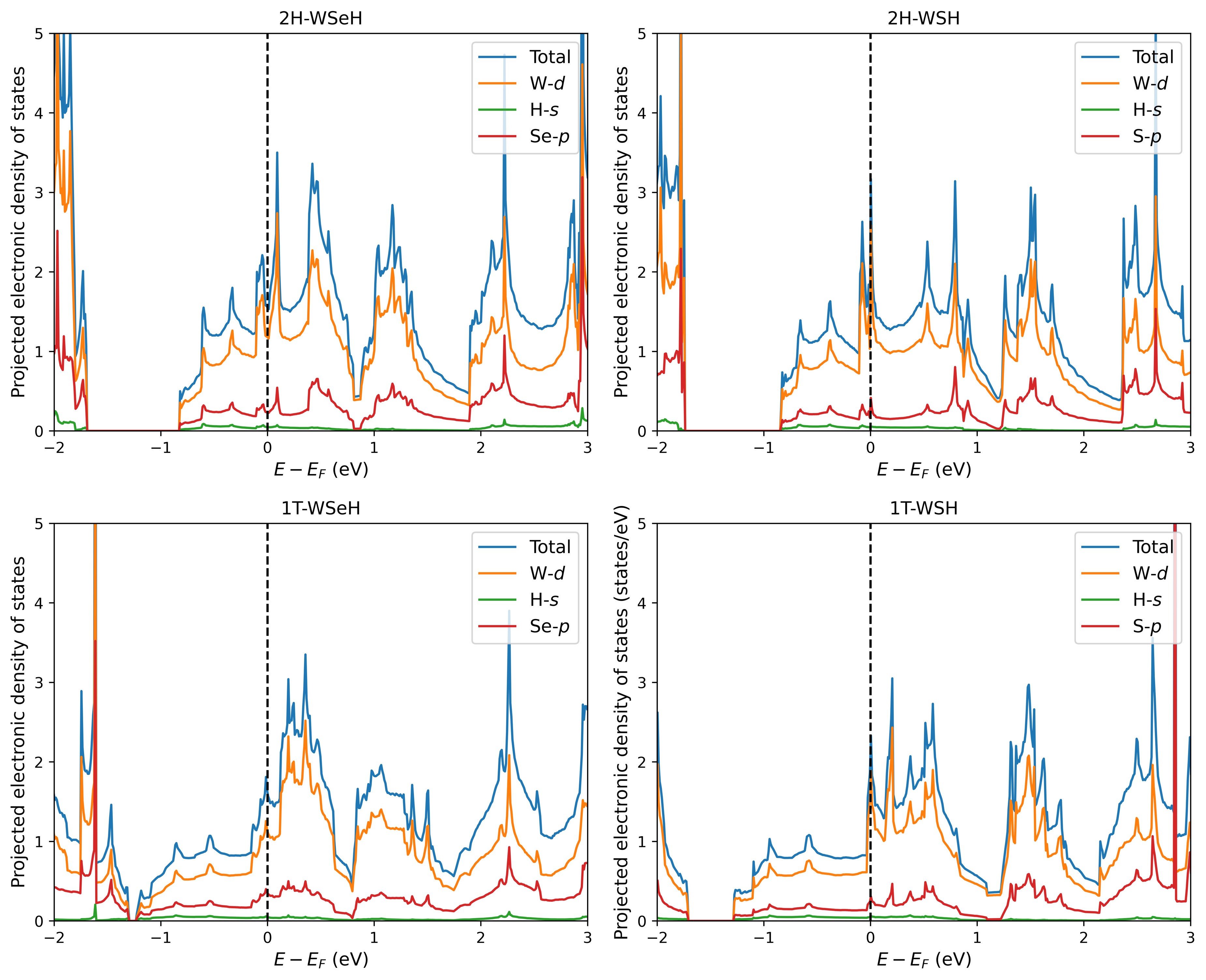}
    \caption{figure shows the projected orbital electronic density of states of 2H-WSeH, 2H-WSH, 1T-WSeH ($\epsilon=-4\%$), and 1T-WSH ($\epsilon=-2\%$). The W-$d$, chalcogenide-$p$ (S-$p$ and Se-$p$), and H-$s$ orbitals are represented by blue, orange, and green lines, respectively.}
    \label{fig:Fig-Appendix-A1}
\end{figure}
The electronic contribution near the Fermi surface is investigated by the projected orbital electronic density of states, as shown in Fig.~\ref{fig:Fig-Appendix-A1}. The density of states reveals a very small fraction of the H-$s$ orbital near the Fermi level, as the bands are predominantly dominated by the W-$d$ orbital, and even smaller contributions are observed from the chalcogenide-$p$ (S-$p$ and Se-$p$) orbital.

\section*{Appendix B: Unstable phonon at $M$ point of 1T structure}
\begin{figure}[H] 
    \centering
    \includegraphics[width=7.5cm]{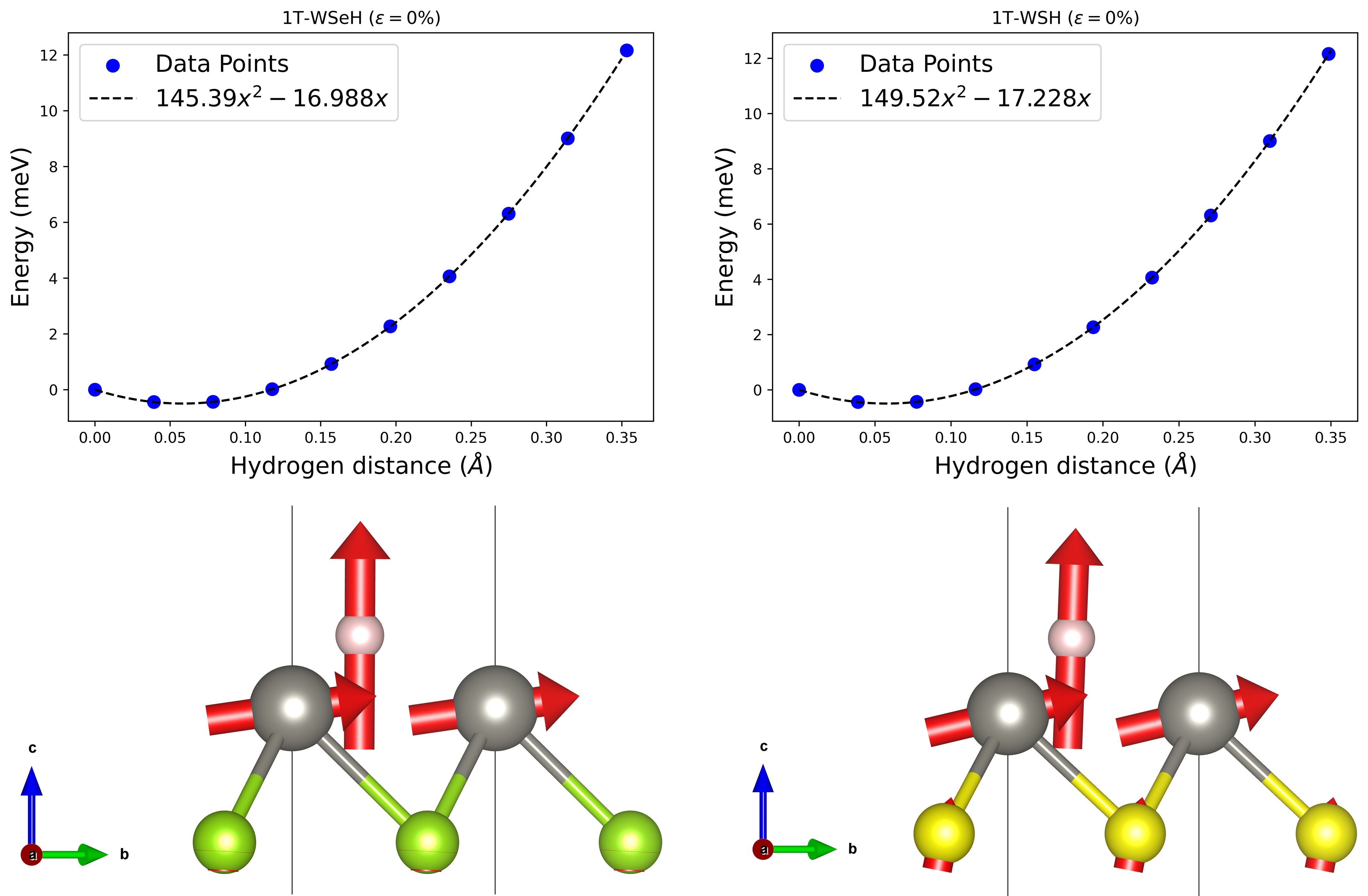}
    \caption{shows the hat-like energy curvature of 1T-WSeH and 1T-WSH indicates the metastable nature of these structures.}
    \label{fig:Fig-Appendix-B1}
\end{figure}
\begin{figure}[H] 
    \centering
    \includegraphics[width=7.5cm]{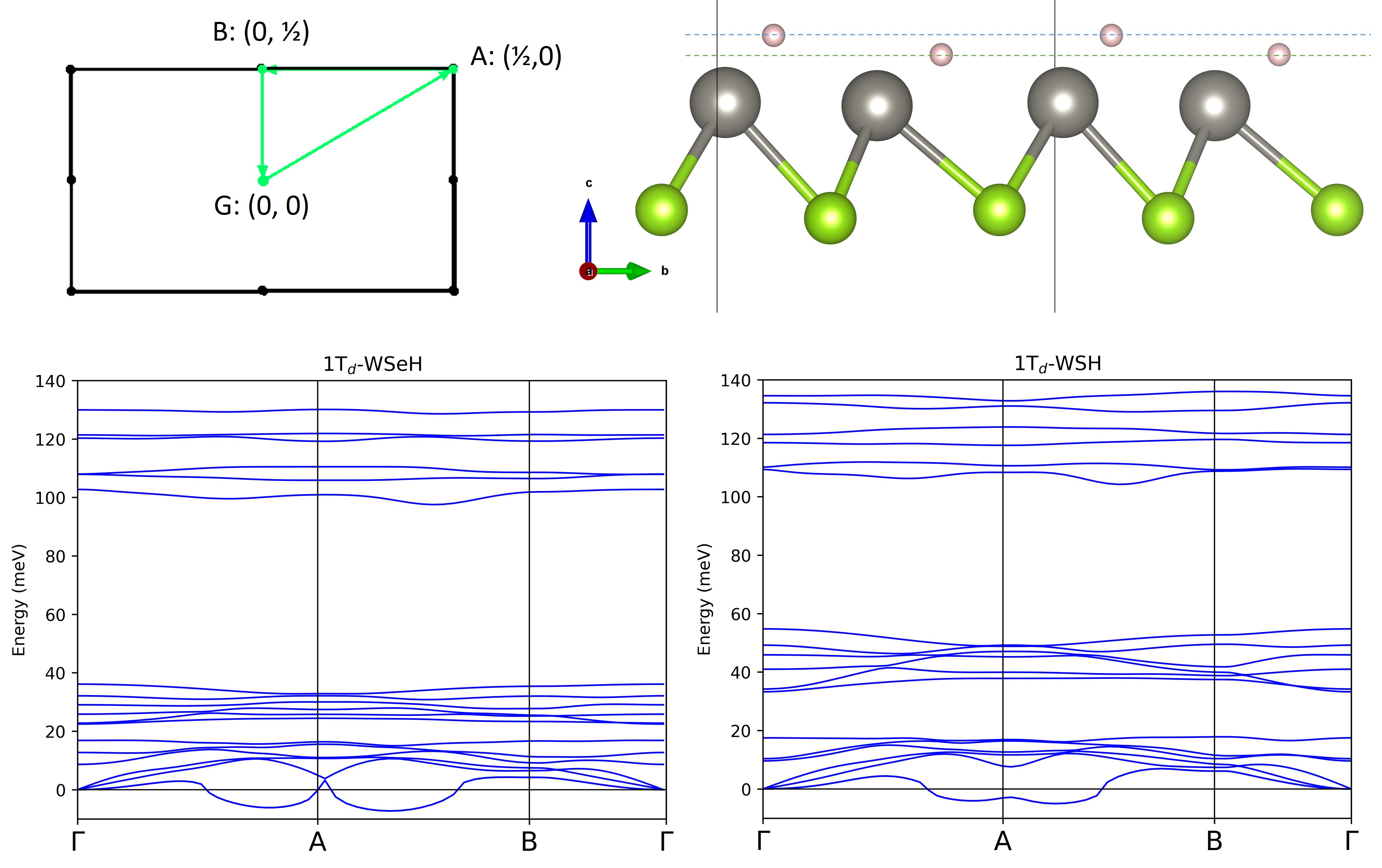}
    \caption{shows paths between $\Gamma$ and zone boundaries $A$ and $B$, the 2$\times$1 distortion of the unit cell, and the phonon dispersions of 1T-WSeH and 1T-WSH.}
    \label{fig:Fig-Appendix-B2}
\end{figure}

From Fig.~\ref{phband-2h1t} (b) and (c), we observe unstable phonons at the $M$ point due to negative phonon energies. Using the frozen-phonon method, the atoms are displaced simultaneously along the eigenvectors at the $M$ point. We determined that these negative phonon energies are the result of a Mexican hat-shaped energy curvature, as shown in Fig.~\ref{fig:Fig-Appendix-B1}. The hat-like energy curvature of 1T-WSeH and 1T-WSH indicates the metastable nature of these structures. However, when strain is applied, it gradually destroys the potential energy curvature of the Mexican hat, as shown in Fig.~\ref{phband-2h1t} (b) and (c) with increasing phonon energies at the $M$ point, and leads to a harmonic energy curvature with positive phonon energy at strains of $\epsilon=-4\%$ and $-2\%$ for 1T-WSeH and 1T-WSH, respectively. Moreover, we analyze the instability at the M point by doubling the unit cell to $2 \times 1$. We observed a distortion in the hydrogen layers, called 1T$_{d}$. However, the phonon dispersion shows a negative frequency around the A point, as shown in Fig.\ref{fig:Fig-Appendix-B2}. Therefore, we applied strain to the system and observed that the distortion vanishes at strains of $-2\%$ and $-4\%$ for 1T$_{d}$-WSH and 1T$_{d}$-WSeH, respectively. On the other hand, we also investigated a $2 \times 2$ distortion of the unit cell, as suggested by the referee. The optimized $2 \times 2$ structures from the final step of molecular dynamics (MD) simulations show a distortion similar to that observed in the $2 \times 1$ distortion. It shows that the energy difference ($E_{1T_d}-E_{2H}$) per atom between the $2 \times 1$ and $2 \times 2$ structures is very similar, at $-51.30$ meV and $-53.52$ meV for WSeH and $-48.46$ meV and $-48.48$ meV for WSH, respectively. We also found that the phonons are not stable for the 1T$_{d}$ $2 \times 2$ distortion in both WSeH and WSH. Furthermore, when applying strains of $-2\%$ and $-4\%$ to the distortion of 1T$_{d}$ $2 \times 2$ of WSeH and WSH, respectively, we found that the distorted hydrogen layer disappears, as in the case of 1T$_{d}$ $2 \times 1$.

\section*{Appendix C: Anharmonicity}
\begin{figure}[H]
    \centering
    \includegraphics[width=7.5cm]{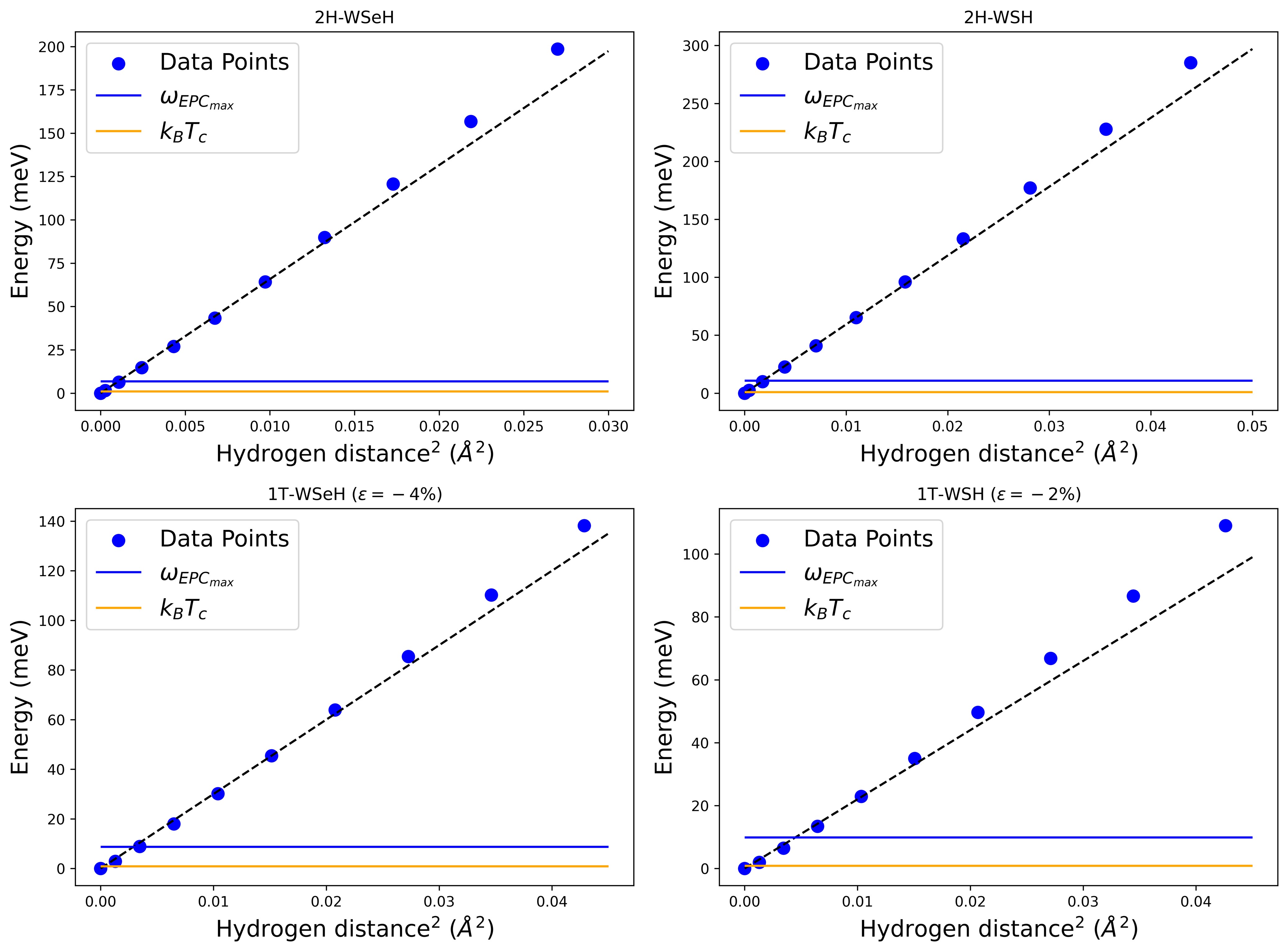}
    \caption{shows the curvature of the adiabatic potential energy surfaces as a function of the square distance of hydrogen atoms of 2H-WSeH, 2H-WSH, 1T-WSeH ($\epsilon=-4\%$), and 1T-WSH ($\epsilon=-2\%$)}
    \label{fig:Fig-Appendix-C1}
\end{figure}
By using the frozen-phonon method, atoms are displaced simultaneously along the eigenvectors of the most strongly coupled phonon mode responsible for the electron-phonon coupling ($\lambda$), specifically arising from the flexural acoustic phonon mode when considering $\lambda_{q\nu}$. For the 2H phase, the most strongly coupled flexural acoustic phonon that we calculated is at $\vec{q} = (0.00000000, 0.48112522)$, while for the 1T phase it is at $\vec{q} = (0.08333333, 0.52923775)$. Fig.~\ref{fig:Fig-Appendix-C1} shows the curvature of the adiabatic potential energy surfaces as a function of the squared distance of hydrogen atoms. The results indicate that the electron-phonon coupling driving the superconducting state is within the harmonic approximation regime.

\section*{Appendix D: The Morel-Anderson pseudopotential ($\mu^*$) dependent T$_c$}
\begin{figure}[H]
    \centering
    \includegraphics[width=6.5cm]{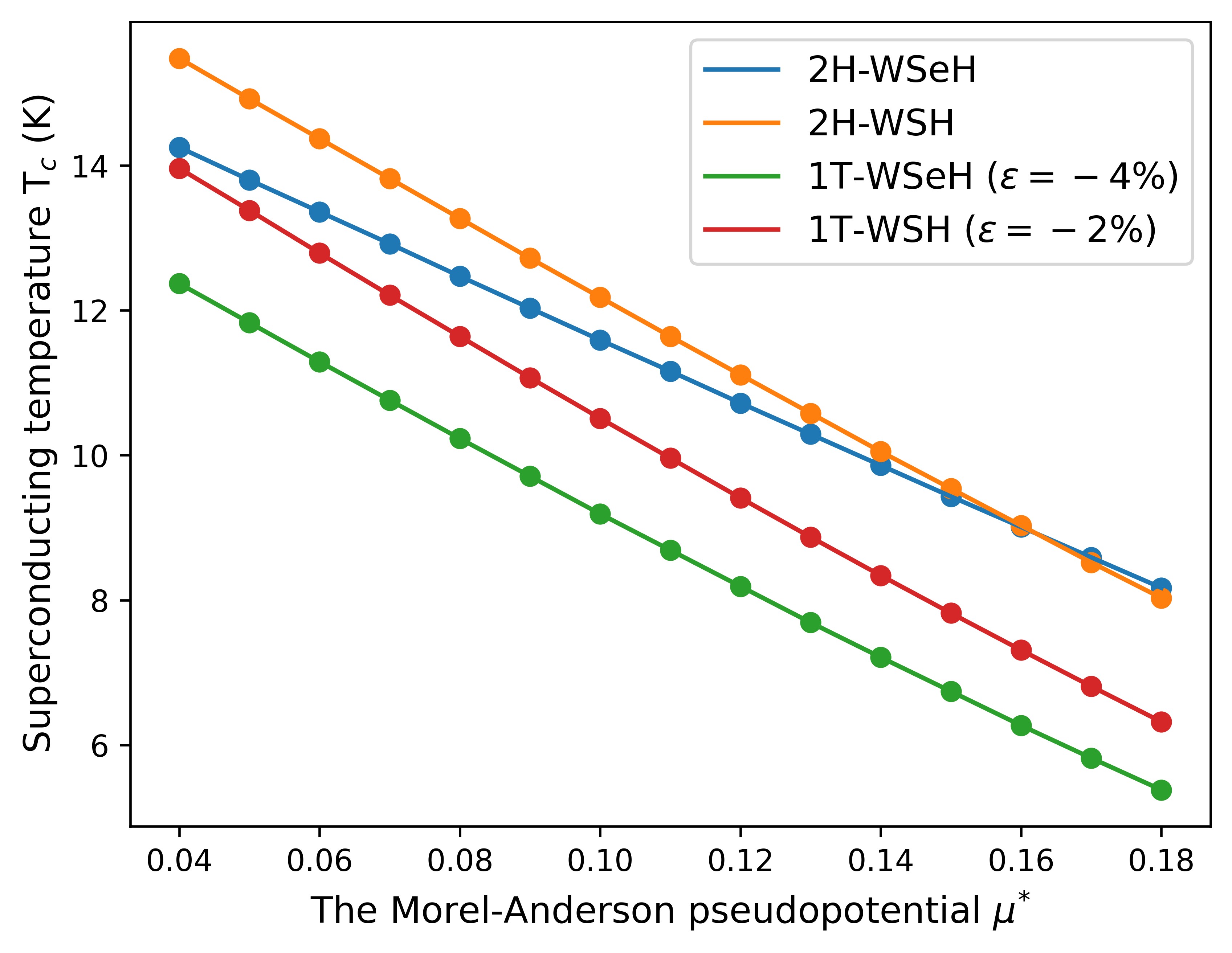}
    \caption{shows the Morel-Anderson pseudopotential ($\mu^*$) dependent T$_c$ of 2H-WSeH, 2H-WSH, 1T-WSeH ($\epsilon=-4\%$), and 1T-WSH ($\epsilon=-2\%$)}
    \label{fig:Fig-Appendix-D1}
\end{figure}
In the main discussion, we compute the superconducting temperature in which the Morel-Anderson pseudopotential was set to $\mu^* = 0.1$ for practical purposes; however, the actual values for these materials could be different, resulting in different values of $T_c$ as shown in Fig.\ref{fig:Fig-Appendix-D1}.



\end{document}